\documentclass[onecolumn]{aastex631}
\usepackage{multirow}

\usepackage{hhline} 

\usepackage{amsmath}
\usepackage{bigints}

\begin{document}

\title{Acceleration of electrons and ions by an ``almost" astrophysical shock in the heliosphere}

\correspondingauthor{Immanuel Christopher Jebaraj}
\email{immanuel.c.jebaraj@gmail.com}

\author[0000-0002-0606-7172]{Immanuel Christopher Jebaraj}
\affil{Department of Physics and Astronomy, University of Turku, 20500 Turku, Finland}

\author[0000-0001-6427-1596]{Oleksiy Agapitov}
\affil{Space Sciences Laboratory, University of California, Berkeley, CA 94720, USA}

\author[0000-0002-6809-6219]{Vladimir Krasnoselskikh}
\affiliation{LPC2E/CNRS, UMR 7328, 45071 Orléans, France}
\affil{Space Sciences Laboratory, University of California, Berkeley, CA 94720, USA}

\author[0000-0002-2238-109X]{Laura Vuorinen}
\affil{Department of Physics and Astronomy, University of Turku, 20500 Turku, Finland}

\author[0000-0003-1236-4787]{Michael Gedalin}
\affiliation{Department of Physics, Ben Gurion University of the Negev, Beer-Sheva 84105, Israel}

\author[0000-0003-2054-6011]{Kyung-Eun Choi}
\affil{Space Sciences Laboratory, University of California, Berkeley, CA 94720, USA}

\author[0000-0001-6590-3479]{Erika Palmerio}
\affil{Predictive Science Inc., San Diego, CA 92121, USA}

\author[0000-0003-3903-4649]{Nina Dresing}
\affil{Department of Physics and Astronomy, University of Turku, 20500 Turku, Finland}

\author[0000-0002-0978-8127]{Christina Cohen}
\affil{California Institute of Technology, Pasadena, CA 91125, USA}

\author{Michael Balikhin}
\affil{University of Sheffield, Sheffield S10 2TN, United Kingdom}

\author[0000-0001-6589-4509]{Athanasios Kouloumvakos}
\affil{The Johns Hopkins University Applied Physics Laboratory, Laurel, MD 20723, USA}

\author[0000-0001-6344-6956]{Nicolas Wijsen}
\affil{Centre for mathematical Plasma Astrophysics, KU Leuven, 3001 Leuven, Belgium}

\author[0000-0002-3298-2067]{Rami Vainio}
\affil{Department of Physics and Astronomy, University of Turku, 20500 Turku, Finland}

\author[0000-0002-4489-8073]{Emilia Kilpua}
\affil{Department of Physics, University of Helsinki, 00014 Helsinki, Finland}

\author[0000-0001-9325-6758]{Alexandr Afanasiev}
\affil{Department of Physics and Astronomy, University of Turku, 20500 Turku, Finland}

\author[0000-0003-1138-652X]{Jaye Verniero}
\affil{Heliophysics Science Division, NASA Goddard Space Flight Center, Greenbelt, MD 20771, USA}

\author[0000-0003-4501-5452]{John Grant Mitchell}
\affil{Heliophysics Science Division, NASA Goddard Space Flight Center, Greenbelt, MD 20771, USA}

\author[0000-0002-0608-8897]{Domenico Trotta}
\affil{The Blackett Laboratory, Department of Physics, Imperial College London, London SW7 2AZ, UK}

\author[0000-0001-7493-2167]{Matthew Hill}
\affil{The Johns Hopkins University Applied Physics Laboratory, Laurel, MD 20723, USA}

\author[0000-0003-2409-3742]{Nour Raouafi}
\affil{The Johns Hopkins University Applied Physics Laboratory, Laurel, MD 20723, USA}

\author[0000-0002-1989-3596]{Stuart D.~Bale}
\affil{Physics Department, University of California, Berkeley, CA 94720, USA}
\affil{Space Sciences Laboratory, University of California, Berkeley, CA 94720, USA}

\begin{abstract}

Collisionless shock waves, ubiquitous in the universe, are crucial for particle acceleration in various astrophysical systems. Currently, the heliosphere is the only natural environment available for their \textit{in situ} study. In this work, we showcase the collective acceleration of electrons and ions by one of the fastest \textit{in situ} shocks ever recorded, observed by the pioneering Parker Solar Probe at only 34.5 million kilometers from the Sun. Our analysis of this unprecedented, near-parallel shock shows electron acceleration up to 6 MeV amidst intense multi-scale electromagnetic wave emissions. We also present evidence of a variable shock structure capable of injecting and accelerating ions from the solar wind to high energies through a self-consistent process. The exceptional capability of the probe's instruments to measure electromagnetic fields in a shock traveling at 1\% the speed of light has enabled us, for the first time, to confirm that the structure of a strong heliospheric shock aligns with theoretical models of strong shocks observed in astrophysical environments. This alignment offers viable avenues for understanding astrophysical shock processes and the acceleration of charged particles.



\end{abstract}

\keywords{}

\section{Introduction} \label{sec:intro}

Collisionless shock waves (CSWs), resulting from converging flows in tenuous plasma, are a fundamental phenomenon in plasma physics \citep[][]{Sagdeev66,Galeev76,Lembege04} capable of heating the plasma and accelerating charged particles \citep[][]{Kennel85,Blandford87}. While ubiquitous in diverse environments, from planetary bow shocks to supernova remnants (SNRs), our comprehension of particle energization (heating and acceleration) at these shocks remains incomplete \citep[][]{Malkov01,Lembege04}. It is mostly the radiation generated by energized electrons that make the detection of CSWs, like those in SNRs, possible \citep[][]{Helder12}. Direct \textit{in situ} examination of astrophysical shocks is currently not possible, leading to a limited understanding of the exact mechanisms behind particle acceleration and ensuing radiation. At present, heliospheric shocks are the only natural CSWs accessible for direct \textit{in situ} study, making them essential for comprehending shock structure, evolution, and particle energization mechanisms. In particular, the shock transition, which marks the finite region between the upstream and downstream flows, plays a crucial role in understanding the redistribution of kinetic energy \citep[][]{Balikhin93,Krasnoselskikh95,Agapitov23}. Consequently, heliospheric shocks, which may in many aspects be similar to young SNR shocks, have become a key focus of research \citep[][]{Terasawa03,Gedalin23}.

In magnetized collisionless shock waves (CSWs), such as those found in supernova remnants (SNRs) and the heliosphere, particle motion is determined by the angle \(\theta_{Bn}\) between the upstream magnetic field and the shock normal. CSWs are classified into quasi-parallel (\(\theta_{Bn} < 45^\circ\)) and quasi-perpendicular (\(\theta_{Bn} > 45^\circ\)) based on this angle. Observational surveys have often found the brightest radio and X-ray synchrotron emissions in regions of SNRs where the shock is quasi-parallel \citep[e.g.][]{Giuffrida22,Vink22}. They are also thought to be the most efficient particle accelerators \citep{Vink20book} and are of particular interest for \textit{in situ} exploration. Quasi-parallel shocks have been understood for nearly 80 years \citep{Moiseev63, Kennel67, Quest88}, but observational efforts have not yet yielded conclusive results \citep{Burgess05, Balikhin23}. Another critical parameter in the study of CSWs is the Alfv\'en Mach number (\(M_{\mathrm{A}}\)), which is directly related to the proportion of flow kinetic energy dissipated at the shock. It is defined as the ratio of the upstream flow speed to the characteristic wave speed in the plasma, known as the Alfv\'en speed (\(v_{\mathrm{A}}\)). Typically, heliospheric shocks exhibit \(M_{\mathrm{A}}\) values less than \(10^2\) \citep[][]{Masters13b}, while SNR shocks can reach \(M_{\mathrm{A}}\) values around \(10^3\) \citep[][]{Vink20book}. A final important parameter is the plasma-\(\beta\), which is the ratio of kinetic plasma pressure to magnetic field pressure. It is generally believed to control the evolutionary characteristics of shocks. Those developing in a low plasma-\(\beta\) regime tend to exhibit large amplitude overshoot magnetic fields \citep{Russell82}.

SNR shocks such as Tycho and Vela Jr.\ expand with velocities around 3000~km~s\(^{-1}\), and are established accelerators of cosmic rays \(>10^{14}\)~eV (100 TeV) by channeling up to \({\sim}10\%\) of their kinetic energy into particle acceleration \citep[][]{Helder09}. A key factor in the acceleration of high-energy particles is the shock's ability to confine particles between converging scattering centers, such as electromagnetic waves, on either side of the CSW. These scattering centers, fundamental to the theory of diffusive shock acceleration (DSA), enable particles to repeatedly cross the shock and gain energy in the process \citep[][]{Krymskii77,Axford77,Blandford78,Bell78}. This energy acquisition process persists until the particles can escape the shock's influence, conventionally dubbed, the duration of particle confinement \citep[][]{Hillas84}. Planetary bow shocks, the strongest heliospheric shocks, are incapable of confining particles long enough to accelerate relativistic particles due to their small size \citep[][]{Hoppe1982,Terasawa11}. 

CSWs driven by solar eruptive events such as coronal mass ejections (CMEs), can occasionally accelerate particles to energies of \(\geq 10^{9}\) eV (GeV) \citep{Reames99}. The most significant acceleration occurs in the relatively denser solar atmosphere when these shocks are fast but not very large, highlighting the importance of their early evolution \citep{Afanasiev18}. This process is somewhat analogous to that of young SNR shocks, which are capable of accelerating particles to \(10^{15}\) eV (PeV) during their rapid evolution within a dense medium \citep{Aharonian19}. A recent statistical study of the strongest CME-driven shocks from the past decade suggests that high-energy electrons are likely accelerated by the shock through a mechanism similar to that affecting ions \citep{Dresing22}. This draws attention to the precise process of shock-induced particle injection and acceleration through DSA, processes that are actively studied \citep[][]{Malkov98,Malkov01}. The complexity of this process is exacerbated by the significant mass disparity between ions and electrons, with protons being 1836 times heavier. This separation of scales necessitates distinct processes for the energization of protons and electrons at their respective scales \citep{Lembege04}.

On March 13, 2023, during its fifteenth perihelion, NASA's Parker Solar Probe (PSP) mission \citep[][]{Fox2016} encountered an extraordinary interplanetary (IP) shock at merely 0.24 astronomical units (a.u.) from the Sun. This CME-driven shock, one of the fastest IP shocks ever recorded \textit{in situ}, was traveling at 1\% the speed of light and exhibited a near-parallel geometry. In this study, we detail the distinct features observed in the transition region of this shock and its impact on particle behavior, enhancing our understanding of strong shocks in both heliospheric and astrophysical contexts. This investigation was enabled by PSP overcoming engineering challenges of entering the Sun's atmosphere, coupled with the exceptional capabilities of its high fidelity instrument suites to measure such phenomena in great detail. Information about PSP's instrument suites used in this study can be found in the Appendix~\ref{App:sec1}.

\section{Results from the analysis}

\subsection{One of the fastest IP shocks observed \textit{in situ} to date}

\begin{figure*}[ht]
    \centerline{
    \includegraphics[width=1\textwidth]{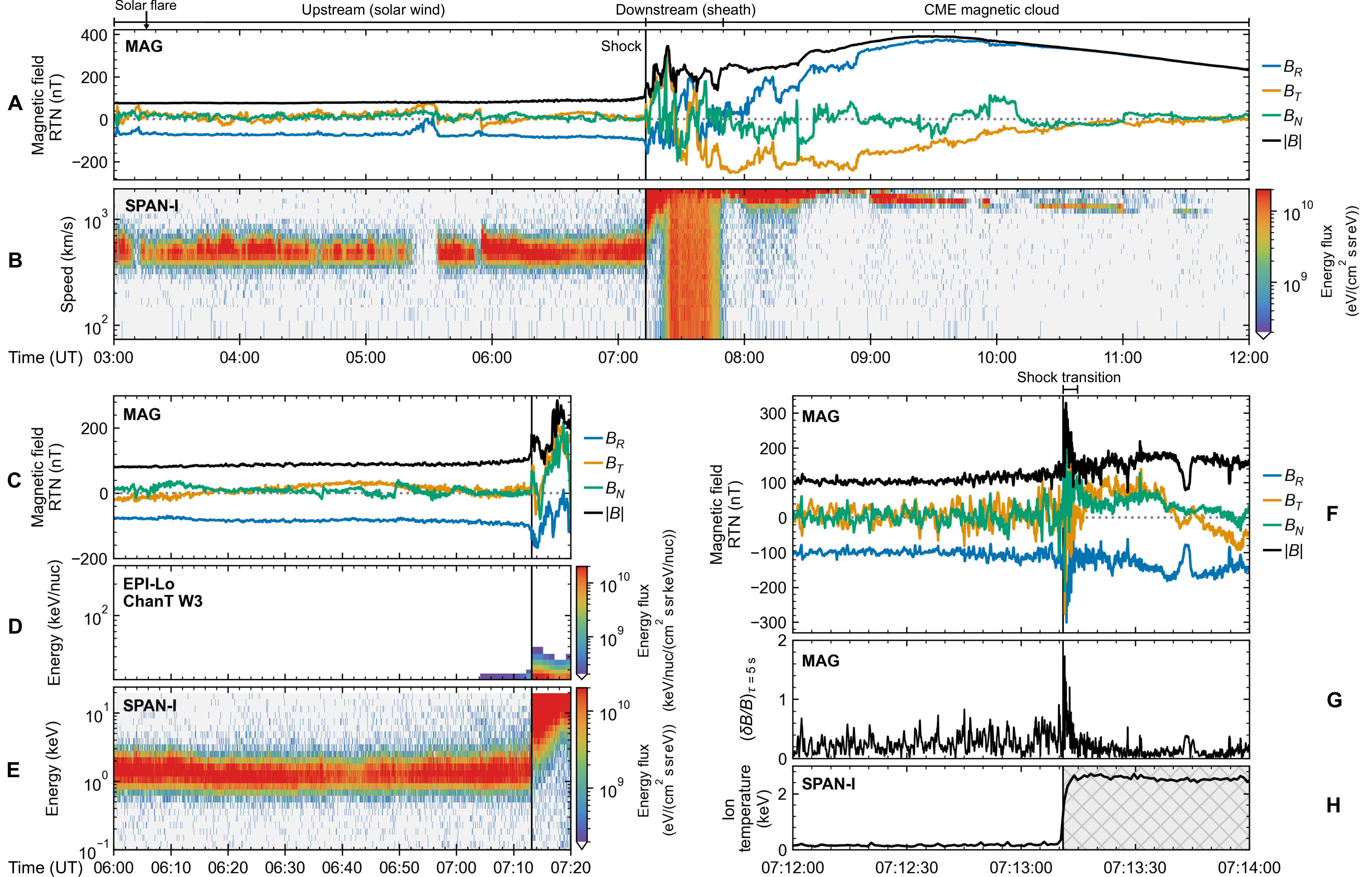}}
    \caption{\textbf{Magnetic field and plasma measurements during the shock encounter.} Panels A and B provide a 9-hour overview of the magnetic field and the one-dimensional proton energy flux, capturing the pre-event solar wind, the \textit{in situ} shock passage, the following sheath region, and the initial portion of the associated CME ejecta. The SPAN-i data downstream of the shock do not accurately represent reality, as the flow is largely deflected away from the detector's field of view. Panels C to E offer an 80-minute close-up leading to the shock arrival, showing variations in the bulk ion distribution as captured by the SPAN-i and EPI-Lo instruments. Finally, Panels F to H focus on a 2-minute snapshot close to the shock transition, emphasizing fluctuations in the total magnetic field as well as the change in proton temperature. The downstream proton temperature is marked with hatching due to significant instrumental uncertainties.}
    \label{fig:fig1}
\end{figure*}

In Fig.~\ref{fig:fig1}(A,B), we present magnetic field and ion measurements during a 9-hour period (03:00 to 12:00~UT) on March 13, 2023. The coordinate system used throughout the manuscript is the inertial RTN (Radial--Tangential--Normal) system, where the radial component R is oriented along the Sun--spacecraft line, the transverse component T is defined to be orthogonal to the rotational axis of the Sun and the radial component, i.e., \(T = \Omega_{\odot} \times R\), while the normal component N completes the orthogonal right-handed triad and, in this case, is aligned with the normal of the ecliptic plane. The two panels provide details on the pre-event conditions and the \textit{in situ} arrival of the shock driven by a large magnetic structure, the CME. The upstream bulk flow speed, measured by SPAN-i, remains consistent throughout the period, averaging \(410 \pm 40\) km s\(^{-1}\) in the 15 minutes prior to the shock's arrival. However, this consistency does not extend to the number density, which is underestimated, as detailed in Appendix \ref{Sec:density_estimation}. Similarly, the upstream magnetic field is averaged over the same period, while the downstream magnetic field is averaged over a subsequent 10-minute period; both are listed in Table \ref{table_1}. 

PSP observed the shock propagating almost exactly radially outward, as indicated by the estimated shock normal \((\hat{\mathbf{n}}_{\mathrm{RTN}}\)), which is also listed in the table. Figure~\ref{fig:fig1}(C--E) provides a zoomed-in view of the shock's arrival at 07:13~UT, marked by a vertical line, and covers the 75 minutes leading up to it. During the shock's passage, the downstream plasma flow exceeded SPAN-i's measurement capabilities, as depicted in Fig.~\ref{fig:fig1}(B,D,E). By combining data from SPAN-i and EPI-Lo and utilizing the principle of mass flux conservation across the shock (detailed in Appendix~\ref{Sec:speed}), we calculate the shock speed to be \(\sim 2800 \pm 300\)~km~s\(^{-1}\) in the spacecraft frame. This classifies it as one of the fastest shocks ever recorded \textit{in situ}, placing it in the same category as the one described in \cite{Russell13}. Additionally, the occurrence of such a strong shock (\(M_{\mathrm{A}} \sim 9.1\pm1.35\)) in a low-\(\beta\) plasma environment (\(\sim 0.16 \pm 0.031\)), along with its near-parallel geometry (\(8^{\circ} \pm 4^{\circ}\)), makes it an exceptionally rare \textit{in situ} observation. A list of the most relevant plasma and shock parameters, along with their uncertainties is presented in Table~\ref{table_1}. The uncertainty for each parameter is subsequently propagated when estimating related parameters, following the general fractional error propagation. For instance, the shock normal is derived as a average from minimum variance analysis (MVA) and the magnetic coplanarity technique (MCT), detailed in Appendix \ref{Sec:shock_normal}. Additionally, unorthodox methodologies employed to estimate plasma density contribute to the uncertainty of the shock parameters, as discussed in Appendix \ref{Sec:density_estimation}.

\begin{table*}
    \centering
    \begin{tabular}{cc}
        Parameter & Data \\
         Shock normal (MVA,MCT), $\hat{\textbf{n}}_{\mathrm{RTN}}$ & [$0.984 \pm 0.01$,$-0.07 \pm 0.036$, $-0.128 \pm 0.096$] \\
         Shock angle $\theta_{Bn}$ ($^{\circ}$) & $8^{\circ} \pm 4^{\circ}$ \\
         Shock speed, $v_{\mathrm{sh}}$ & $2800 \pm 300$ km s$^{-1}$ \\
         Upstream magnetic field, $B_{\mathrm{u}}$ & $90\pm5$ nT \\
         Downstream magnetic field, $B_{\mathrm{d}}$ & $140\pm7$ nT \\
         Overshoot magnetic field, $B_{\mathrm{max}}$ & $330$ nT \\
         Magnetic compression, $r_B$ & $1.55\pm0.1$ \\
         Upstream electron density (RFS), $n_{\mathrm{u}}$ & $46.5 \pm 15$ cm$^{-3}$ \\
         Downstream electron density (RFS), $n_{\mathrm{d}}$ & $184 \pm 35$ cm$^{-3}$  \\
         Density compression, $r_{\mathrm{gas}}$ & \(4\) \\
         Upstream ion inertial length, $\lambda_{\mathrm{i}}$ & $32.5 \pm 3$ km \\
         Upstream bulk flow speed, $v_{\mathrm{u}}$ & $410\pm40$ km s$^{-1}$ \\
         Downstream bulk flow speed, $v_{\mathrm{d}}$ & $2200 \pm 200$ km s$^{-1}$ \\
         Upstream Alfv\'en speed, $v_{\mathrm{A}}$ & $268 \pm 20$ km s$^{-1}$ \\
         Upstream ion beta, $\beta_{\mathrm{u}}$ & $0.16 \pm 0.031$ \\
         Alfv\'en Mach number, $M_{\mathrm{A}}$ & $9.1\pm 1.35$ \\
    \end{tabular}
    \caption{List of relevant plasma and shock parameters associated with the March 13, 2023 event.}
    \label{table_1}
\end{table*}

A notable characteristic of quasi-parallel shocks is the patchwork of complex magnetic structures making up an extended transition region, typically in the order of 100s of ion-inertial lengths \citep[\(\lambda_\mathrm{i}\), the characteristic distance over which ions respond to electromagnetic changes.][]{Burgess05}. This contrasts with quasi-perpendicular shocks, known for their sharp transition in the form of a ramp and overshoot \citep{Krasnoselskikh13}. This complexity of quasi-parallel shocks is illustrated in Fig.~\ref{fig:fig1}(F--H), which display the magnetic field, its fluctuations, and the ion temperature over a 2-minute interval (07:12--07:14 UT). As we approach the shock transition, marked by the vertical line, there is a noticeable increase in the amplitude of magnetic field fluctuations (\(\delta B/B\)), indicative of intense upstream wave activity. The data also show that the magnitude of the magnetic field intensifies before the transition, likely as a result of shock mediation by energetic particles. This amplification of the magnetic field and the intensification of fluctuations is also predicted by theoretical models of strictly parallel shocks \citep[e.g.][]{Wang22}. The shock transition is characterized by a \(\delta B/B \geq 0.5\), preceding the ramp-overshoot region, which exhibits a \(\delta B/B \geq 2\). Figure~\ref{fig:fig1}H displays a sudden rise in ion temperature at the transition, consistent with the expected behavior for magnetohydrodynamic (MHD) shocks \citep{Sagdeev66}. While it is evident that the temperature sharply increases at the transition, the SPAN-i measurements do not accurately reflect downstream conditions.

\subsection{Electron acceleration to ultra-relativistic energies}

\begin{figure*}[ht!]
    \centerline{
    \includegraphics[width=1\textwidth]{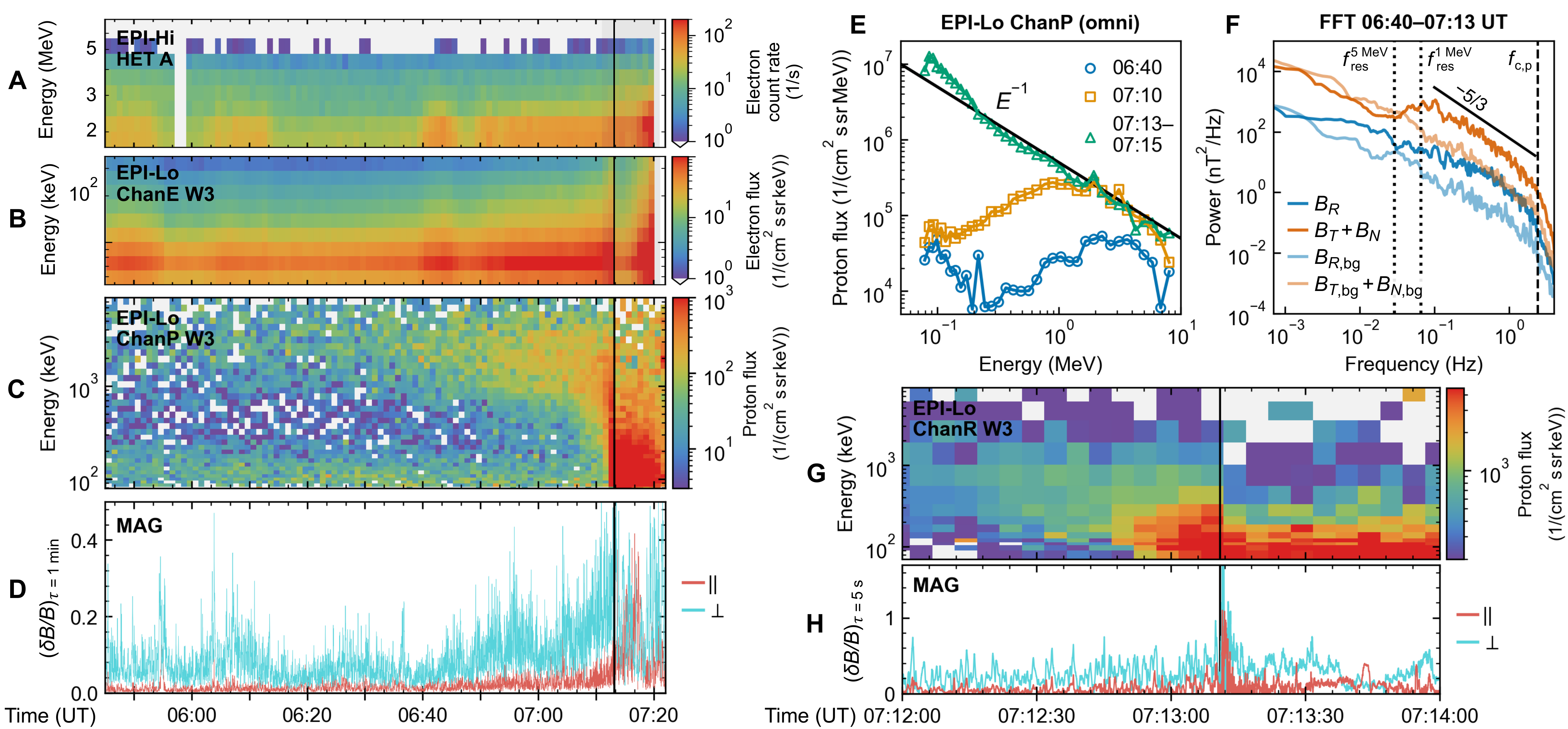}}
    \caption{\textbf{Energetic particles and the evolution of the wave foreshock.} Panels A--D show the time--energy plots for high-energy electron counts detected by EPI-Hi, low-energy electron flux from EPI-Lo, proton flux from EPI-Lo, and the amplitude of transversal (\(\perp\)) and parallel (\(\parallel\)) magnetic fluctuations. In panel E, we present the proton intensity spectra for three selected times (06:40, 07:10, and at the shock, averaged as 07:13--07:15 UT) with a \(E^{-1}\) power-law shown for reference, which indicates the changes in energetic particle distribution. Panel F shows the Fourier power spectra of \(B_R\) (estimating \(B_{\parallel}\) power) and the sum of \(B_T\) and \(B_N\) (estimating \(B_{\perp}\) power) in the foreshock and the background. Variations in the proton flux and magnetic fluctuations over a two-minute period surrounding the shock transition are shown in panels G and H.}
    \label{fig:fig2}
\end{figure*}

A particularly notable and unprecedented observation concerning this strong quasi-parallel shock is the local acceleration of electrons, with energies ranging from tens of keV (sub-relativistic) to 6~MeV (ultra-relativistic). Observations of the local acceleration of relativistic electrons (300~keV to 3~MeV) is exceptionally rare in IP shocks \citep{Nasrin23,Jebaraj23} and is even less frequent at planetary bow shocks, regardless of their strength \citep{Masters13}. The occurrence of ultra-relativistic electrons at these shocks has never been documented before. The rarity of local electron acceleration at IP shocks can be attributed to a combination of scale-related challenges and a lack of high-fidelity instrumentation. Physical challenges, such as the energization of electrons from a thermal core, require specific processes that affect electrons at scales close to their gyrofrequency \((\omega_\mathrm{ce}\)). These processes are related to fundamental issues of energy redistribution in CSWs \citep{Balikhin93, Schwartz11}. Additionally, their acceleration also necessitates the presence of oblique waves across a wide range of frequencies. Figure~\ref{fig:fig2}(A,B) presents energetic electron observations during a 90-minute period surrounding the shock's arrival. Figure~\ref{fig:fig2}A displays the counts per second (s\(^{-1}\)) of relativistic electrons (\(>\)1~MeV) throughout this interval, while Fig.~\ref{fig:fig2}B provides differential energy fluxes of lower-energy electrons, ranging from 50 to 350 keV. Notably, there is a significant increase in electron fluxes and counts across a wide range of energy when approaching the shock (between 06:40 and 07:13~UT), thus illustrating its impact on electron acceleration. 

The observed range of electron energies, spanning sub-relativistic to ultra-relativistic, implies the presence of multi-scale electromagnetic structures in and around the shock transition, affecting electrons across all these energies. Fig.~\ref{fig:fig2}B shows that while sub-relativistic electrons are significantly affected by the shock transition, relativistic and ultra-relativistic electrons seem largely unaffected. This distinction arises from the increasing radius of gyration (Larmor radius) of electrons with higher kinetic energy. The upstream data reveal fluctuations in electron fluxes at distinct times, specifically at 06:00 UT and 06:40 UT, suggesting localized upstream phenomena. Nonetheless, the limited resolution of electron observations restricts a more detailed analysis of these variations.

\subsection{Self-consistent injection and acceleration of solar wind ions}

In Fig.~\ref{fig:fig2}C, we showcase the ion fluxes measured by EPI-Lo's sunward-pointing detector (W3) at the shock, and up to 90 minutes prior, to investigate the motion of energetic ions. Measurements from all EPI-Lo wedges are presented in Figs.~\ref{fig:figs4} and \ref{fig:figs5} of Appendix~\ref{appendix_Epiplots}. A striking observation is the intense population of ions streaming ahead of the shock, persisting for the entire 90-minute duration, peaking between 06:40 and 07:13~UT. Figure~\ref{fig:fig2}D illustrates the fluctuations in the magnetic field (\(\delta B/B\)) on a characteristic time scale of \(\tau=1\) minute, corresponding to a frequency of approximately 0.02 Hz. This frequency roughly matches the resonance frequency for the lowest energy of the streaming ions further from the shock, namely \(\sim\)1~MeV. The fluctuations are categorized into transversal (\(\delta B_{\perp}/B\)) and parallel (\(\delta B_{\parallel}/B\)) components, with the former being significantly larger than the latter within the same timeframe. The technical details of these analyses can be found in Appendix \ref{appendix_e}. 

Another intriguing feature is the observed absence of flux at energies below \({\sim} 1\) MeV, before 07:05~UT. This absence of flux extends to higher energies further from the shock, especially prior to 06:40~UT. In Fig.~\ref{fig:fig2}E, the omni-directional intensity spectra of protons are shown for three distinct time periods: at the shock (07:13--07:15~UT), near-upstream (07:10~UT), and far-upstream (06:40~UT). These spectra highlight a marked absence of flux near 1~MeV, followed by a significant flux increase towards 10~MeV, forming what can be described as a spectral roll-over at low energies. The significant absence of low-energy particles away from the shock may be due to the fact that only high-energy particles diffuse farther from the shock. Approaching the shock however, the spectrum evolves into a single power law with a spectral index of \(E^{-1}\) beyond \(\sim\)250 keV. Below approximately 250 keV, the spectral index becomes steeper, approaching \(E^{-2}\). However, we refrain from interpreting this finding, as it is potentially erroneous to perform a spectral fit over a range that does not extend across at least an order of magnitude.

Figure~\ref{fig:fig2}F presents the Fourier power spectra for \(B_{\parallel}\) and \(B_{\perp}\) during a 33-minute interval upstream (06:40--07:13~UT) and compares it to the pre-event background (01:30--02:07~UT). For a detailed description of the spectral analysis, please refer to Appendix~\ref{appendix_FFT}. Generally, upstream power for both \(B_{\parallel}\) and \(B_{\perp}\) exceed pre-event levels by more than an order of magnitude at frequencies above the resonance frequency for particles of energy \({\sim}1\) MeV (\(f_{\mathrm{res}}^{\mathrm{1 MeV}}\)). Below this frequency, we observe little difference between the upstream and pre-event power of \(B_{\perp}\). In contrast, there is an increase in \(B_{\parallel}\) relative to the pre-event power above \(f_{\mathrm{res}}^{\mathrm{5 MeV}}\). However, it is important to note that frequencies at or below \(10^{-2}\) Hz are subject to significant uncertainties due to various factors, as detailed in the Appendix \ref{appendix_FFT}.

Zooming into the shock ramp, we note an uptick in both the amplitude and intensity of these fluctuations. In Fig.~\ref{fig:fig2}(G,H), we show the energetic protons and magnetic field fluctuations within a 2-minute window (07:12--07:14~UT) around the ramp. Here, we detect an intense proton population with energies between 100 and \(\sim\)250 keV which correspond to the energy range where the spectra becomes steeper in Fig. \ref{fig:fig2}E. This occurs concurrently with high-intensity, large-amplitude \(\delta B_{\perp}/B \geq 0.5\) fluctuations. These fluctuations are estimated at \(\tau=5\) seconds, corresponding to resonant frequencies above 0.2 Hz, which match those of protons in the 100s of keV range. The \(\delta B_{\parallel}/B\) also show an increase as we approach the ramp, especially after 07:13~UT. These observations concur with prior studies on the challenges of particle confinement at quasi-parallel shocks leading to an extensive wave foreshock \citep{Kennel86}. It also supports theoretical predictions that wave power declines with increasing distance from the shock owing to the reduced particle fluxes \citep{Bell78,Vainio07}. 

Finally, the continuous detection of ions with energies less than 100~keV to 10~MeV near the shock is noteworthy. Figure~\ref{fig:fig2}C highlights data from the W3 detector of EPI-Lo, but all EPI-Lo detectors registered ions above 1 MeV for about 90 minutes before the shock arrival. In contrast, \(\leq\)300~keV protons were primarily observed near the shock transition. It is likely that these are either reflected or gyrating ions as previously suggested by \cite{Kennel81} and thus deviate from the \(E^{-1}\) law below \(\sim 250\) keV. Distinguishing between the two components would require higher-cadence particle measurements. However, in the case of near-specular reflection, the energy of the ions would primarily peak at twice the speed (four times the energy) of the upstream bulk flow (in the shock rest frame), as most of the ions are reflected only once \citep{Gedalin16}. Considering the quasi-parallel geometry of the shock, these ions can diffuse ahead of the shock and participate in DSA, thereby gaining significantly higher energies. An \(E^{-1}\) scaling law supports this process occurring in a lossless manner, meaning that the protons (ions) are confined in the vicinity of the shock without being lost. Conversely, this could also suggest that they are replenished at the same rate as they are lost, underscoring the efficiency of the shock in particle acceleration.

\begin{figure*}[ht]
    \centerline{
    \includegraphics[width=1\textwidth]{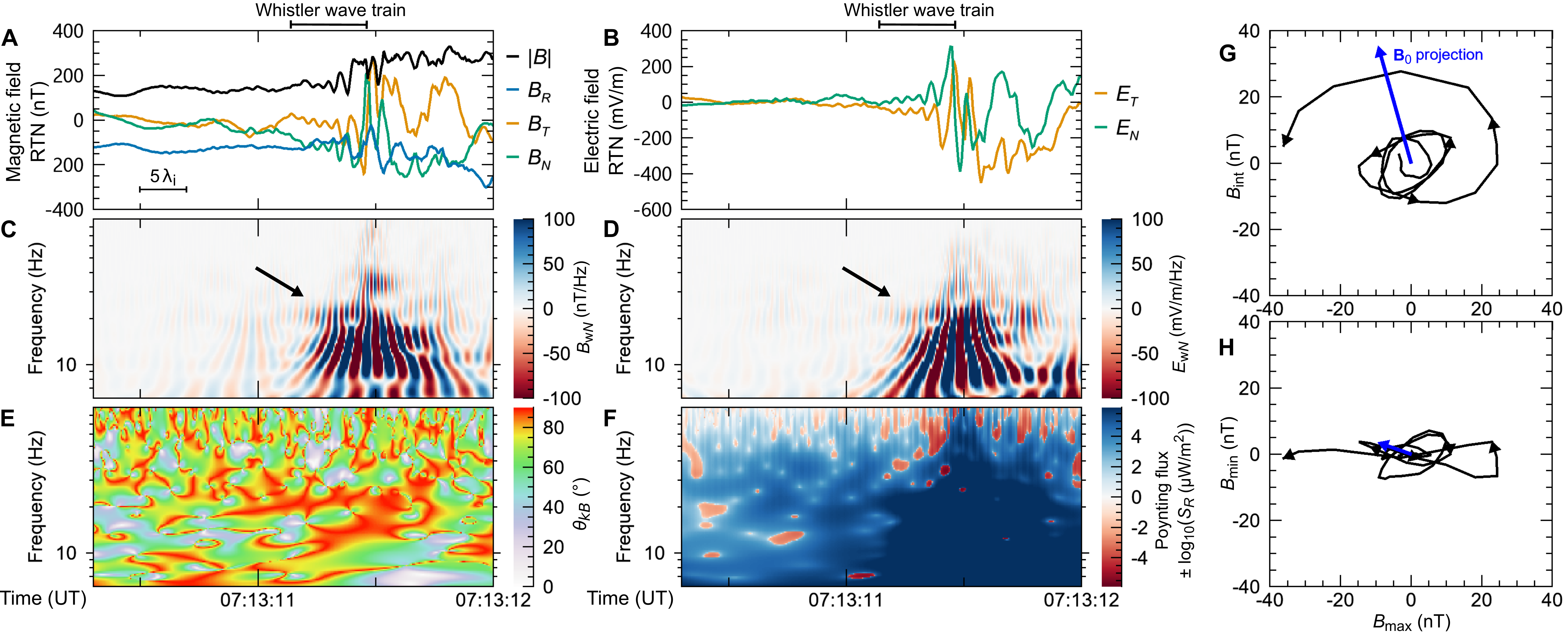}}
    \caption{\textbf{Analysis of the waves at the shock transition.} Panels A and B show the magnetic and electric field components. The duration of the precursor whistlers (between 07:13:11.10--07:13:11.45) is annotated on top. Panel A also indicates the approximate spatial scale of the upstream in units of \(\lambda_\mathrm{i}\) estimated in the plasma frame. Panels C and D show the Morlet wavelet spectrum of the magnetic and electric fields. The $\theta_{kB}$ angle between the wave vector $\textbf{k}$ and the magnetic field $\textbf{B}$ is shown in panel E. In panel F, we show the Poynting flux estimated in the spacecraft frame $\textbf{S}^{sc}$. Finally, the hodographs of precursor whistler waves in the maximum vs intermediate variance plane and the maximum vs minimum variance planes are presented in panels G and H, respectively. In both panels, the blue arrow denotes the mean field \(B_0\).}
    \label{fig:fig3}
\end{figure*}

\subsection{Structure of the quasi-parallel shock transition}

In contrast to the infinitesimal boundary between two flows marking a shock in MHD, the shock transition is where processes crucial for regulating the distribution of kinetic energy occur, such as dispersion, dissipation, and non-linear steepening \citep{Krasnoselskikh85b,Krasnoselskikh2002}. In Fig.~\ref{fig:fig3}, we present a comprehensive analysis of the shock transition, with further details available in Appendix~\ref{appendix_f}. Figure~\ref{fig:fig3}A illustrates the magnetic field components and magnitude over a 1.75-second interval (07:13:10.25--07:13:12.00), complemented by Fig.~\ref{fig:fig3}B, which depicts the electric field.

Figure~\ref{fig:fig3}(C,D) depict the dynamic spectra of magnetic and electric field perturbations, respectively, derived from the Morlet wavelet transform. These panels emphasize the shock ramp and its associated precursor wave train, notably within the 20--30~Hz frequency range (indicated by arrows) which is higher than the ion gyrofrequency (\(\omega_\mathrm{ci} \sim 1.4\)~Hz). The wave train observed here consists of circularly polarized waves, as seen in Fig.~\ref{fig:fig3}G, the plane containing the intermediate and maximum variance axis. Consequently, they also exhibit high planarity in the plane containing the maximum and minimum variance axis as shown in Fig.~\ref{fig:fig3}H. Such characteristics indicate that these are electromagnetic whistler mode waves, which are occasionally observed ahead of the shock ramp and are associated with fundamental shock processes \citep{Sundkvist12}. These waves are right-hand circularly polarized in the spacecraft frame and occur between the ion and electron gyrofrequencies (\(\omega_\mathrm{ci} < \omega < \omega_\mathrm{ce}\)). As the wave number (\(k\)) increases and they approach \(\omega_\mathrm{ce}\), these waves can also increase in amplitude and may steepen.

The wave normal angle, depicted in Fig.~\ref{fig:fig3}E (\(\theta_{kB}\), the angle between the wave normal and the background magnetic field direction), indicates that both the precursors and the ramp are quasi-perpendicular to the magnetic field. Their oblique propagation is an essential factor to be a standing wave in the shock frame. Figure~\ref{fig:fig3}F illustrates the radial component of the Poynting flux, estimated in the spacecraft frame, indicating an anti-sunward propagation direction of the observed wave activity near the shock. The technical material related to the analysis of the precursor whistler waves can be found in Appendix~\ref{appendix_whistlers}. From this analysis, the estimated phase speed (\(v_{\mathrm{ph}}\)) is approximately 400~km~s\(^{-1}\). Given that \(\theta_{kB}\) is approximately \(75^\circ\), the radial component is calculated to be around \(2600 \pm 200\)~km~s\(^{-1}\). This analysis of the precursor whistler wave train lends credibility to the estimated shock speed. 

More importantly, we demonstrate that the ramp and precursors display a quasi-perpendicular geometry. This finding is in line with theoretical predictions \citep{Gedalin98,Gedalin15}, which suggest that a quasi-parallel shock becomes oblique, even quasi-perpendicular, due to the increase in transversal magnetic field amplitude in the ramp-overshoot region. However, prior to our study, observational support for this was limited \citep[][]{Balikhin23}. We have clearly demonstrated that the enhancement of the transversal magnetic field at the ramp-overshoot leads to a quasi-perpendicular geometry. The degree of enhancement is particularly noteworthy given that the shock develops in a low plasma-\(\beta\) environment, where large ramp-overshoot structures have previously not even been observed in either oblique or quasi-perpendicular obstacle shocks \citep[][]{Russell82}.

\section{Discussion}

\begin{figure*}[t!]
    \centering
    \includegraphics[width=1\textwidth]{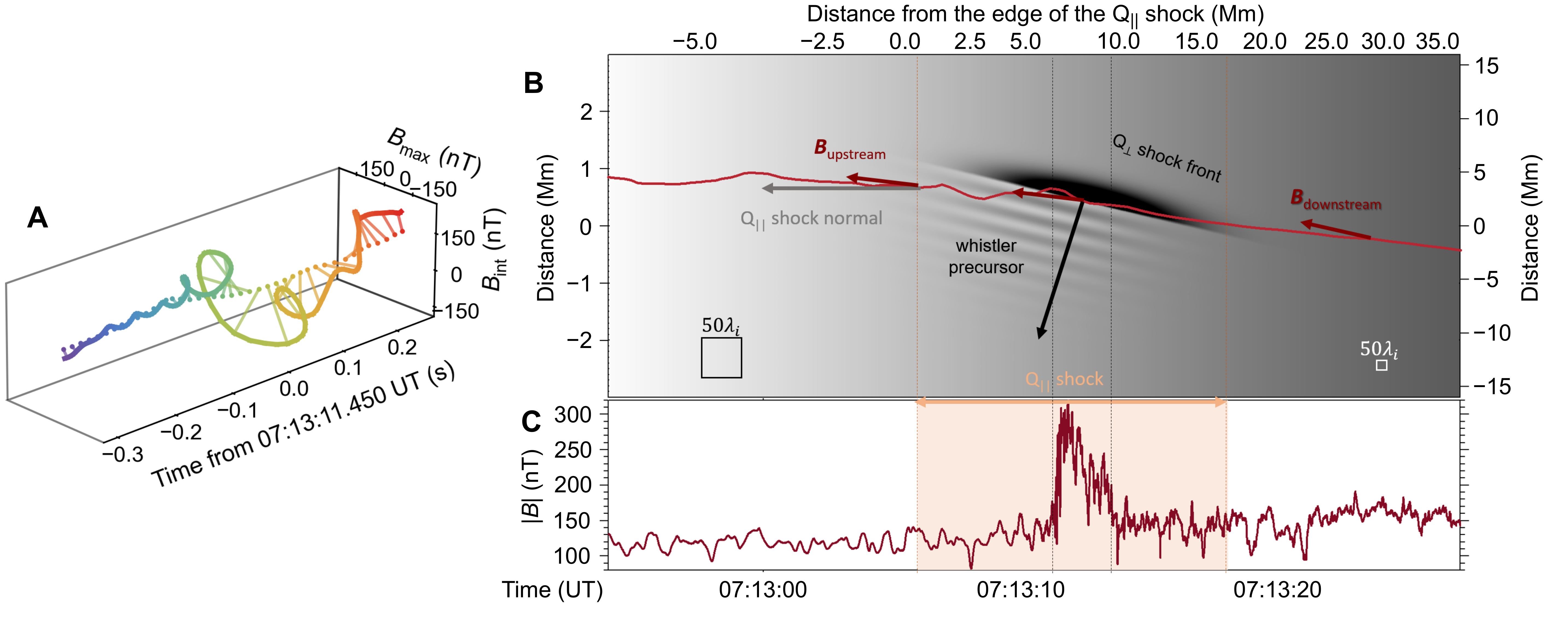}
    \caption{\textbf{Transition region of the quasi-parallel shock.} Panel A displays the time evolution (depicted by the color blue to red) of the transversal magnetic field components. The dotted line through the middle represents the mean field, \(B_0\). Panel B illustrates a two-dimensional schematic of the shock transition in the plasma frame. The spatial scales are in \(\lambda_\mathrm{i}\) (represented with 50\(\lambda_\mathrm{i} \times\)50\(\lambda_\mathrm{i}\) squares) and are estimated for the upstream and downstream separated at the edge of the quasi-parallel shock (Q\(_{\parallel}\)). The grayscale represents the change in \(\textbf{B}\) across the shock. The direction of \(\textbf{B}\) and the shock normal are also marked by arrows. The dark stripes in the middle represent the quasi-perpendicular structure at the ramp and the whistler precursors. Panel C presents the magnetic field magnitude on which the schematic is based upon.}
    \label{fig:fig4}
\end{figure*}


Strong collisionless shock waves, spanning a wide range of Mach numbers in the heliosphere, have been extensively studied to enhance our understanding of particle acceleration in astrophysical shocks, such as those in SNRs. However, in contrast to SNR shocks, even the strongest heliospheric shocks (planetary bow shocks) are ineffective accelerators of relativistic particles \citep{Hoppe1982}. This inefficiency is largely attributed to their smaller size and the weak magnetization of the ambient medium, resulting in inadequate particle confinement \citep{Terasawa11}. Conversely, interplanetary traveling shocks are significantly larger but rarely as strong as planetary bow shocks. Remote observations have occasionally identified CME-driven shocks possessing strengths comparable to planetary bow shocks and speeds similar to some SNR shocks \citep{Yurchyshyn2005}. Particle acceleration by these shocks is anticipated to be highly efficient \citep{Afanasiev18}. Measurements from PSP enable us to investigate these shocks \textit{in situ}, unprecedentedly close to the Sun. This proximity significantly narrows the gap between our understanding of heliospheric shocks and SNRs, allowing for an in-depth study of efficient acceleration mechanisms.

In this study, we have presented extraordinary observations of a strong (\(M_{\mathrm{A}} \sim 9.1 \pm 1.35\)), fast (\(v_{\mathrm{sh}} \sim 2800 \pm 300\)~km~s\(^{-1}\)), and near-parallel (\(\theta_{Bn} \sim 8^{\circ} \pm 4^{\circ}\)) traveling interplanetary shock. This shock, one of the fastest IP traveling shocks observed \textit{in situ} to date, exhibits a magnetic structure that aligns with theoretical models. Figure~\ref{fig:fig4}A illustrates the clear rotation of the transverse magnetic field components at the shock transition, not obscured by nonlinear structures formed in the foreshock. This alignment with theoretical predictions is essential for making reasonable comparisons with SNR shocks.

As expected from such a prominent shock, it proved to be an efficient accelerator of particles. A notable observation is the local acceleration of electrons across a wide range of energies by such a near-parallel shock, often disregarded as potential accelerators of energetic electrons \citep{Masters13}. Typically, in the MHD context of a near-parallel shock, the magnetic field amplification is minimal and smooth, lacking fine structures, which means the electrons are barely affected by it. Electron acceleration presents a significant challenge due to the substantial scale difference from ions, as highlighted by the mass ratio. In the case of quasi-perpendicular shocks, electron acceleration is believed to occur via simple adiabatic reflection due to the large magnetic gradient. However, only a small number of particles participate due to the steep energy requirements for injection \citep{Jebaraj23}. Conversely, in quasi-parallel shocks, where there is no single, large magnetic gradient, electron acceleration may require the presence of high-frequency oblique wave activity. These waves can then induce either adiabatic or non-adiabatic behavior in electrons. The presence of such waves is illustrated in the schematic of the shock transition presented in Fig.~\ref{fig:fig4}B. 


The schematic, based on the magnetic field data in Fig.~\ref{fig:fig4}C, details various spatial scales and directions for clarity. It demonstrates that at scales where the shock transition is considered a simple MHD discontinuity, it is quasi-parallel. At smaller scales, within the shock's transition region, we identify whistler precursors and the ramp-overshoot structure. This area is marked by quasi-perpendicular electromagnetic waves, instrumental in the injection, trapping, and eventual acceleration of electrons to relativistic energies. Such processes have been supported by numerical simulations \citep{Riquelme2011}. Once electrons attain sufficient energy and their Larmor radii increase adequately, they can interact with ion-driven waves. Thus, accurately quantifying these distinct spatial scales is critical, as it influences the redistribution of flow kinetic energy into processes like electron heating and acceleration \citep{Balikhin93,Schwartz11}.


Finally, we have demonstrated for the first time that the quasi-perpendicular ramp-overshoot region reflects ions, whereas the quasi-parallel upstream region allows their escape. The latter inability of the quasi-parallel shock to effectively confine particles at the ramp, results in the generation of ion-scale waves in the upstream region. These phenomena collectively create a continuous process at the shock transition, essential for particle injection and sustained acceleration, resulting in an uninterrupted ion population. We have observed this continual presence of ions with energies ranging from less than 100 keV to 10 MeV. We found the concurrent occurrence of \({>}1\) MeV streaming ions and intense transverse waves up to 90 minutes prior to shock arrival, suggesting active diffusive shock acceleration. This is supported by the low-energy roll-over observed in the upstream energy spectra, which is unstable leading to the growth of ion-driven waves \citep{Shapiro98}. Conversely, the confinement of low-energy ions near the shock leads to a consistent \(E^{-1}\) energy spectrum, in line with DSA predictions (with no losses) for a shock with \(r_{\mathrm{gas}} = 4\).


In this study, we used \textit{in situ} evidence to demonstrate that a strong quasi-parallel shock can self-consistently accelerate both ions and electrons from the background medium to high energies. Such as self-consistent mechanism is enabled by the amplification of both the magnitude of the mean magnetic field and the level of multi-scale fluctuations relative to it. This is compelling even from an elementary prediction that a shock must either be very large or highly magnetized to effectively confine high energy particles \citep{Hillas84,Terasawa11}. Our findings naturally lead to the conclusion that strong shocks close to the Sun are highly effective at particle confinement by virtue of being highly magnetized. This makes them comparable to much larger shocks with lesser magnetization. These rare observations were made possible by the pioneering capabilities of the Parker Solar Probe and its unprecedented proximity to the Sun. If the physics of cosmic shocks, like those in SNR blast waves, adhere to the fundamental principles of CSWs, then understanding the evolution of the shock transition is crucial for facilitating the acceleration of charged particles.


\subsection*{Acknowledgments} 
\textbf{Author contributions:} The study was initiated by ICJ and conceptualized together with OVA, VVK, and MG. The methodologies for data analysis was provided by ICJ, OVA, VVK, LV, MG, and EP. Data analysis was performed by ICJ, OVA, LV, MG, KEC, EP, and AK. The IS\(\odot\)IS data was processed by JGM, LV, CMSC, ND, and AK. The SPAN-i data was prepared by JV and analyzed by ICJ, OVA, LV, and KEC. The results were interpreted by ICJ, OVA, VVK, LV, MG, and RV. The data was visualized by ICJ, OVA, LV, and KEC. ICJ wrote the manuscript and prepared the final draft. The final draft was prepared taking into account extensive comments from OVA, VVK, MG, MB, ND, RV, EKJ, and SDB. All authors have agreed to the results presented in the manuscript.

\textbf{Funding information:}
The Parker Solar Probe spacecraft was designed, built, and is now operated by the Johns Hopkins Applied Physics Laboratory as part of NASA’s Living with a Star (LWS) program (contract NNN06AA01C). Support from the LWS management and technical team has played a critical role in the success of the Parker Solar Probe mission.
The authors express their gratitude to all the instrument teams for their work in processing and publishing the publicly available data from the Parker Solar Probe. All the data used in this study are from the Parker Solar Probe spacecraft. The data used in this study are available at the NASA Space Physics Data Facility (SPDF), https://spdf.gsfc.nasa.gov.
This research was supported by the International Space Science Institute (ISSI) in Bern through ISSI International Team project No.~575, ``\textit{Collisionless Shock as a Self-Regulatory System}''.
ICJ, ND, and LV are grateful for support by the Academy of Finland (SHOCKSEE, grant No.~346902), and the European Union’s Horizon 2020 research and innovation program under grant agreement No.~101004159 (SERPENTINE).
OVA was partially supported by NSF grant number 1914670, NASA’s Living with a Star (LWS) program (contract 80NSSC20K0218), and NASA grants contracts 80NNSC19K0848, 80NSSC22K0433, 80NSSC22K0522. OVA and VVK were supported by NASA grants 80NSSC20K0697 and 80NSSC21K1770.
VVK also acknowledges financial support from CNES through grants ``Parker Solar Probe'' and ``Solar Orbiter''.
LV acknowledges the financial support of the University of Turku Graduate School. 
EP acknowledges support from NASA's Parker Solar Probe Guest Investigators (PSP-GI; grant no.~80NSSC22K0349) program.
NW acknowledges support from the Research Foundation -- Flanders (FWO-Vlaanderen, fellowship no.~1184319N). 
AK acknowledges financial support from NASA NNN06AA01C (SO-SIS Phase-E, PSP EPI-Lo) contract.
JLV acknowledges support from NASA PSP-GI grant 80NSSC23K0208.
The FIELDS experiment was developed and is operated under NASA contract NNN06AA01C.
\\


\appendix


\section{Experimental details} \label{App:sec1}

\subsection{Electromagnetic fields (FIELDS)}

Our study primarily utilizes full electromagnetic fields measured by the FIELDS instrument suite on board the PSP spacecraft \citep{Bale16}. The electric field measurements are made using the electric fields instrument (EFI) consisting of two pairs of dipole electric field antennas oriented in the TN plane and extending beyond the PSP heat shield, and a fifth antenna located behind the heat shield on the instrument boom; the location of antenna V5 in the wake of PSP means the R component is susceptible to detrimental interference by the wake electric field and cannot be reliably interpreted \citep{Bale16}. Two three-component flux-gate magnetometers (MAG) measure the magnetic field from DC to approximately 60 Hz during aphelion and up to 293 vector measurements per second during 2 to 4 days around perihelion. The latter is used in the present study. 

Additionally, we incorporate high-frequency electric field measurements from both the high-frequency and low-frequency receivers (HFR \& LFR) of the Radio Frequency Spectrometer \citep[RFS;][]{Pulupa2017}. The RFS includes four electric antennas and measures over a wide frequency range, spanning from 1 kHz to 20 MHz at a 3.5 second cadence during the science phase. A particularly relevant application of these measurements in our study is in determining the electron plasma frequency, \(\omega_{\mathrm{pe}}\).

\subsection{Solar Wind Electrons, Alphas, and Protons (SWEAP)} \label{Sec:SWEAP}

Considering that shock waves manifest as discontinuities in both electromagnetic fields and plasma, our study employed the Solar Wind Electrons, Alphas, and Protons (SWEAP) instrument suite \citep{Kasper16}, specifically using the ion electrostatic analyzer referred to as the Solar Probe ANalyzer for Ions (SPAN-i) \citep{Livi22}. SPAN-i measures the 3D velocity distributions of solar wind ions with an additional time-of-flight component to distinguish between protons and heavier ion species. For this event, we used the L3 sf00 data product corresponding to uncontaminated protons at a $~$3.5s cadence in the energy range $~$20 eV to $\sim$20 keV. However, it is important to note a limitation: due to the placement of the instrument on the ram side of PSP, its field of view (FOV) is obstructed by the Thermal Protection System (TPS), which means only partial plasma moments are available. Significant deflections in the solar wind flow (characterized by the angle between the magnetic field vector \(\textbf{B}\) and the plasma flow velocity vector \(\textbf{v}\), denoted as \(\theta_{Bv}\)), with respect to the spacecraft frame, may result in parts of the bulk distribution being undetected by the instrument. When the proton velocity distribution function (VDF) sufficiently departs from the FOV of SPAN-i, the temperature measurement is largely overestimated since the instrument will only capture the wings of the distribution. 

An example of this deflection is evident at 05:30~UT in Fig.~1B, where the VDF remains unrecorded, as deflections in the +T direction are most susceptible to measurement reliability.Additionally, during the shock crossing period, the proton VDF not only sufficiently deflected from the FOV, but also exceeded the energy range of SPAN-i. Consequently, the downstream plasma measurements become entirely artificial and fail to reflect the actual conditions. Regarding the temperature at the time of the shock's arrival, the proton VDF's tails are exceptionally broad, resulting in an anomalously high measured temperature. During the period of time when the upstream velocity was obtained, however, the VDF was sufficiently in the FOV, as verified by plotting the VDFs from the L2 sf00 data product and confirming that the majority of the bulk flow is unobstructed (not shown). However, parts of the VDF other than the core may be unaccounted for when estimating the number density resulting in discrepancies. 

\subsection{Integrated Science Investigation of the Sun (IS$\odot$IS)}

To analyze the energetic particle populations, in particular electrons and protons, we used the Integrated Science Investigation of the Sun, IS\(\odot\)IS instrument suite \citep{McComas16}. It measures energetic particles from $\sim$20 keV to over 100~MeV/nuc with two Energetic Particle Instruments (EPI), EPI-Lo \citep{Hill17} and EPI-Hi \citep{Wiedenbeck17}. EPI-Lo is a time-of-flight mass spectrometer with 80 separate apertures to provide a ${\sim}2\pi$-wide FOV. We use a few different data products from EPI-Lo, namely, the ChanP (triple coincidence protons), ChanR (high time cadence triple coincidence proton measurements), ChanT (time-of-flight-only ion measurements), and ChanE (used primarily to measure electrons) channels. EPI-Hi includes two low energy telescopes (LETs) measuring ions from $\sim$1 to 20 MeV/nuc; one double-ended with a sunward facing aperture, LETA, and anti-sunward aperture, LETB, and a single-ended telescope, LETC, pointing orthogonally to the LETA instrument axis. The higher energies (${>}10$~MeV/nuc) are captured in EPI-Hi by a double-ended high energy telescope (HET) with one side, HETA, pointed sunward and the other, HETB, anti-sunward. The sunward and anti-sunward designations are based on the assumption of the nominal magnetic field line.

\section{Plasma parameter estimations}

\subsection{Estimation of the downstream flow speed}

The downstream distribution of the bulk plasma flow is beyond the measurement capabilities of the SPAN-i instrument for E15, which corresponds to approximately 20 keV or 2000~km~s$^{-1}$. Consequently, it was not feasible to directly obtain plasma parameters such as density and velocity at the moment of the shock crossing, parameters that are crucial for understanding the shock wave's characteristics.

To gauge the extent to which the downstream bulk flow was shifted, we utilized data from the EPI-Lo ChanT, which extends down to 30 keV. This is illustrated in Fig.~\ref{fig:fig1}D of the main text, where part of the distribution is observable at 30 keV. This observation strengthens the assertion that the bulk flow's kinetic energy exceeded SPAN-i's measurement range and was not merely deflected away from its sensor. However, it is a complex task to match a thermal distribution across two different instruments. Therefore, we infer that the center of the distribution lies somewhere between 20 and 30 keV. Since these measurements represent the kinetic energy (\(E_{\mathrm{ke}}\)) of the bulk flow, we estimated the downstream velocity (\(v_{\mathrm{d}}\)) using the formula \(v_{\mathrm{d}} = \sqrt{2E_{\mathrm{ke}}/m_{\mathrm{p}}}\), where \(m_{\mathrm{p}}\) denotes the proton mass. This calculation yields a downstream velocity of approximately \(2200\,\pm\,200\)~km~s$^{-1}$. The \(\pm 200\)~km~s$^{-1}$ uncertainty is due to the large energy bin width (20 to 30~keV) that was considered. Comparatively, the upstream bulk flow velocity (\(v_{\mathrm{u}}\)), measured by the spacecraft, was around \(410\pm40\)~km~s$^{-1}$ in the 2-minute interval preceding the shock's arrival. While it is possible to get an approximation for the magnitude, \(v_{\mathrm{d}}\) of the downstream flow vector, it is not possible to obtain the direction of the flow. 

\subsection{Estimation of electron density} \label{Sec:density_estimation}

To verify the plasma density measurements from SPAN-i, we employed the electron plasma frequency measured by the FIELDS/RFS electric field antennas as a proxy. Under usual conditions, the RFS antennas determine electron density from quasi-thermal noise \citep{Meyer-Vernet17}. However, this approach was not feasible during the observed period because the Debye length (\(\lambda_\mathrm{De}\)) was larger than the effective length of the antennas \citep{Moncuquet20}. In such cases, electrostatic waves, and especially electron beam-driven Langmuir waves typically resonating near the electron plasma frequency (\(\omega_{\mathrm{pe}}\)) can be used. In cases where electron sound waves are present, a result of electron cyclotron instabilities, the frequency can be downshifted to approximately 30--60\% of \(\omega_{\mathrm{pe}}\) (i.e., \(0.4-0.7 \omega_{\mathrm{pe}}\)) \citep{Lobzin05}. Therefore, the highest frequency waves detected in our observations are presumed to be Langmuir waves.

When observed with a dipole antenna pair like the FIELDS/RFS (namely \(V_{1}-V_{2}\) and \(V_{3}-V_{4}\)), electrostatic waves should exhibit high phase coherence \citep{Krasnoselskikh11}. We constructed a phase coherence spectrum to identify such coherent signals both upstream and downstream of the shock. Coherence can be estimated as,

\[\gamma^2 = \frac{(\tilde{X}_0 \tilde{X}_1^*)(\tilde{X}_0 \tilde{X}_1^*)^*}{(\tilde{X}_0 \tilde{X}_0^*)(\tilde{X}_1 \tilde{X}_1^*)}.
\]

In the above equations, \(\tilde{X}_0 \tilde{X}_0^*\) is the auto spectrum from the (Channel 0) \(\vec{V}_{12} = \vec{V}_{1} - \vec{V}_{2}\) dipole, and \(\tilde{X}_1 \tilde{X}_1^*\) is the auto spectrum from the (Channel 1) \(\vec{V}_{34} = \vec{V}_{3} - \vec{V}_{4}\) dipole. \(\tilde{X}_0 \tilde{X}_1^*\) is the cross spectrum between the two channels.

\begin{figure*}[ht]
    \centering
    \includegraphics[width=1\textwidth]{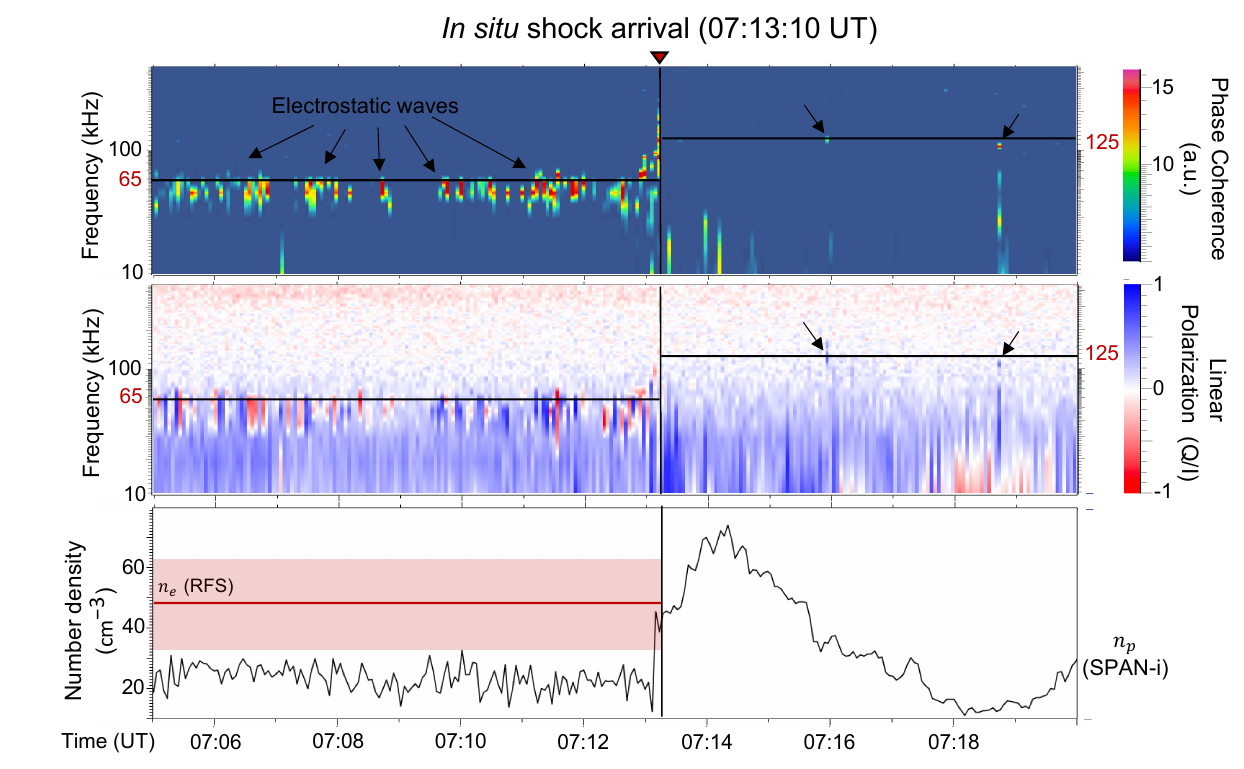}
    \caption{\textbf{\textit{In situ} arrival of the shock as observed through high-frequency electric field data from the FIELDS/RFS.} The top panel displays the coherence spectra, while the middle panel presents the normalized linear polarization, Stokes Q/I. The estimated upstream and downstream plasma frequencies (\(\omega_{\mathrm{pe}}\)) are indicated by horizontal black lines, with arrows highlighting the observed electrostatic waves. The bottom panel juxtaposes the estimated electron density (\(n_{\mathrm{e}}\)) – with the range of uncertainty depicted as a red shaded area – against the ion density measured by SPAN-i.}
    \label{fig:figs1}
\end{figure*}

The top panel of Fig.~\ref{fig:figs1} displays this coherence spectrum, clearly distinguishing upstream electrostatic waves near the \(60 \pm 10\) kHz frequency channels. We focused exclusively on highly coherent signals, enabling distinct identification of the waves at the shock transition and the upstream. Non-stationary processes and general noise typically exhibit near-zero phase coherency. Intriguingly, two coherent downstream signals were also detected, marked by arrows, at frequencies of \(125 \pm 10\) kHz. A notable capability of the FIELDS/RFS receivers is their measurement of full Stokes parameters, providing robust verification of the wave types observed.

Langmuir waves, being highly field-aligned, are strongly linearly polarized. Linear polarization is quantified by the Stokes Q parameter, normalized with the total intensity (Stokes I), expressed as Q/I. Both Stokes Q and I are estimated as,

\[I = \tilde{X}_0 \tilde{X}_0^* + \tilde{X}_1 \tilde{X}_1^*\]

\[Q = \tilde{X}_0 \tilde{X}_0^* - \tilde{X}_1 \tilde{X}_1^*\]

The second panel of Fig.~\ref{fig:figs1} presents these strongly linearly polarized electrostatic waves upstream, with the coherent downstream signals also showing a relatively high degree of linear polarization.

During the upstream interval from 07:06 to 07:12~UT, the highest frequency recorded was \(65 \pm 10\) kHz. This uncertainty is due to the frequency channel's \(4.5\%\) bandwidth and the variation in the highest frequency wave observed during this period. For the downstream interval, we utilized the frequencies of the two strongly coherent and linearly polarized signals, \(125 \pm 10\) kHz.

To estimate the electron plasma density using Langmuir waves, we can use the linear dispersion relation for Langmuir waves in the spacecraft frame:
\[
\omega = \omega_{\mathrm{pe}} + \frac{3v_{\mathrm{th}}^{2}k^{2}}{4\omega_{\mathrm{pe}}} + kV_{\mathrm{sw}} \cos \theta_{kB},
\]
where \(k\) is the wavenumber, \(v_{\mathrm{th}} = \sqrt{2k_{B}T_\mathrm{e} / m_\mathrm{e}}\) represents the electron thermal speed, and \(\theta_{kB}\) is the angle between the solar wind flow and the wave vector \(\textbf{k}\). The equation includes a final term accounting for the Doppler shift caused by the plasma flow past the spacecraft. These two factors may cause \(\omega\) to deviate from \(\omega_{\mathrm{pe}}\): the increase in frequency due to the thermal speed and the Doppler shift. It is noteworthy to mention that the Doppler shift may be both positive and negative and accounts for the largest contribution to the observed frequency. The range of \( k \) was estimated by \cite{Graham21}, who found \( 0.11 \leq k \lambda_\mathrm{De} \leq 0.25 \) for typical parameters found in the solar wind. For instance, they considered an electron temperature \( T_\mathrm{e} \sim 15 \pm 5 \) eV, similar to the values obtained by \cite{Liu23} using the PSP measurements, \( T_\mathrm{e} \sim 20 \pm 5 \) eV . The Doppler shift accounts for a solar wind speed \( V_\mathrm{sw} \sim 400 \) km/s, which is similar to the \( v_\mathrm{u} = V_\mathrm{sw} \) measured upstream of the shock. \cite{Graham21} also considered linearly polarized Langmuir waves, which are field-aligned. By using these parameters in the linear dispersion relation, we estimated the Langmuir wave frequency to range from \( 0.95\omega_\mathrm{pe} \) to \( 1.15\omega_\mathrm{pe} \). As expected, the variation in \(\omega\) is largely attributable to the Doppler effect. It is worth noting that this estimation might change when considering a more accurate but complex dispersion relation, such as the quasi-electrostatic slow extraordinary mode wave presented in \cite{Larosa21}. The longitudinal component of the wave, indicated by the ratio \(\frac{\omega_\mathrm{ce}^2}{\omega_\mathrm{pe}^2}\), becomes prominent when \(k\) is on the order of this ratio. However, for the conditions presented here, where \(\frac{\omega_\mathrm{ce}^2}{\omega_\mathrm{pe}^2} \sim 10^{-3} - 10^{-2}\) assuming \(\omega \approx \omega_\mathrm{pe}\), its effect is minimal.

The \(\omega_{\mathrm{pe}}\) is of particular interest, as it is directly related to the electron plasma density (\(n_{\mathrm{e}}\)) which is given by the formula:

\[ \omega_{\mathrm{pe}} = \sqrt{\frac{n_{\mathrm{e}} e^2}{\epsilon_0 m_{\mathrm{e}}}} \]

where, \(e\) is the elementary charge, \(\epsilon_0\) is the vacuum permittivity, and \(m_{\mathrm{e}}\) is the electron mass. The \(n_{\mathrm{e}}\) can be estimated by rearranging the formula for \(\omega_\mathrm{pe}\):


\[ n_{\mathrm{e}} = \frac{\omega_{\mathrm{pe}}^2 \epsilon_0 m_{\mathrm{e}}}{e^2} \]

Given the range of \(\omega\) with respect to \(\omega_\mathrm{pe}\), we can directly estimate a range of values for \(\omega_\mathrm{pe}\). We estimate the \(\omega_\mathrm{pe}\) corresponding to \(\omega = 65\pm10\) kHz is \(\omega_\mathrm{pe} = 61.5\pm10\, \text{kHz}\), and for \(\omega \sim 125\pm10\) kHz we estimate \(\omega_\mathrm{pe} = 121\pm12 \, \text{kHz}\). Next, by applying the measured values of \(\omega_{\mathrm{pe}}\), we can estimate \(n_\mathrm{e}\). The upstream density using \(\omega_\mathrm{pe} = 61.5\pm10\, \text{kHz}\) results in an estimated average electron plasma density of \(46.5 \pm 15\) cm\(^{-3}\). However, data from SPAN-i, as shown in the last panel of Fig.~\ref{fig:figs1}, indicate an average proton density of approximately \(22\) cm\(^{-3}\), which is a factor of two smaller than our estimate. While, \(\omega_\mathrm{pe} = 121\pm12 \, \text{kHz}\) corresponds to an average electron density of \(184 \pm 35\) cm\(^{-3}\). We may consider the SPAN-i measurement of number density to be a lower limit in our estimations. 


\section{Shock parameter estimations} \label{Sec:shock_speed}

\subsection{Shock normal} \label{Sec:shock_normal}

Identifying the shock wave's normal direction necessitates pinpointing its transition. To achieve this, we implemented a moving average technique with a 30-second sliding window to filter out high-frequency signals and isolate the magnetic gradient. We observed an increase in the mean field 30 seconds prior to the transition, indicated by the principal shock jump at 07:13~UT. For the estimation of the normal, we employed two methods: minimum variance analysis \citep[MVA;][]{Sonnerup98}, and the magnetic coplanarity theorem \citep[MCT;][]{Colburn66}. Due to uncertainties and the absence of necessary plasma measurements such as velocity, we could not apply other methods such as velocity coplanarity and mixed-mode techniques, usually regarded as more robust \citep{Paschmann00}.

For the MVA technique, applied when a spacecraft encounters a transition layer like a shock front, the relevant equations are:

\[ \sum_{\nu=1}^{3} M_{\mu\nu} n_\nu = \lambda_\mu n_\mu \]

Here, \( M_{\mu\nu} \) equals \( \langle \mathbf{B}_\mu \mathbf{B}_\nu \rangle - \langle \mathbf{B}_\mu \rangle \langle \mathbf{B}_\nu \rangle \), and \( \hat{\textbf{n}}_{\mathrm{MVA}} \) aligns with the smallest eigenvalue, \( \lambda_{\mathrm{min}} \). We assess reliability using the ratio \( \lambda_{\mathrm{int}} / \lambda_{\mathrm{max}} \) and \( B_{\mathrm{n}} / B \), with \( \lambda_{\mathrm{int}} / \lambda_{\mathrm{max}} \geq 2 \) and \( B_{\mathrm{n}} / B \leq 0.3 \) indicating a well-defined shock normal.

We selected the time window for MVA to span from 07:03 to 07:23~UT, i.e.\ centered on the shock ramp at 07:13~UT. Within this interval, the eigenvalue ratio between the intermediate and minimum variance directions, \( \lambda_{\mathrm{int}}/\lambda_{\mathrm{min}} \), was approximately 7.5, suggesting a good distinction---the three eigenvalues are $\lambda_{\mathrm{max}}=8298$, $\lambda_{\mathrm{int}}=3070$, and $\lambda_{\mathrm{min}}=408$. The normal, indicated by the direction of the minimum variance eigenvector, was determined to be \([0.976, -0.052, -0.216]\) in the inertial-RTN coordinate system. However, the magnetic field ratio \( B_{\mathrm{n}}/B \approx 0.66 \) suggested that the normal direction was not well-defined, even after accounting for the high-frequency components of the field fluctuations. For a well-defined normal direction, it is expected that \( B_{\mathrm{n}}/B < 0.1 \). However, in the case of quasi-parallel shocks, where \(B_\mathrm{n}\) is on the order of \(B\), the ratio of eigenvalues (\( \lambda_{\mathrm{int}}/\lambda_{\mathrm{min}} \)) serves as a far better indicator of quality \citep{Paschmann00}.

To corroborate these findings, we applied the MCT, the only other feasible method using solely magnetic field data. The MCT calculates the shock normal \( \hat{\textbf{n}}_{\mathrm{MCT}} \), assuming \( \mathbf{B}_{\mathrm{u}} \), \( \mathbf{B}_{\mathrm{d}} \), and \( \hat{\mathbf{n}} \) are coplanar:

\[ \hat{\mathbf{n}}_{\mathrm{MCT}} = \pm \frac{(\mathbf{B}_{\mathrm{d}} - \mathbf{B}_{\mathrm{u}}) \times (\mathbf{B}_{\mathrm{d}} \times \mathbf{B}_{\mathrm{u}})}{\|(\mathbf{B}_{\mathrm{d}} - \mathbf{B}_{\mathrm{u}}) \times (\mathbf{B}_{\mathrm{d}} \times \mathbf{B}_{\mathrm{u}})\|} \]

We set the upstream and downstream windows for this analysis from 07:04 to 07:12~UT and from 07:15 to 07:23~UT, respectively. The choice of the interval is motivated by previous statistical studies of \textit{in situ} shocks \citep{Kilpua15}. The estimated shock normal \(\hat{\textbf{n}}\) was found to be \([0.995, -0.101, -0.026]\) in the RTN coordinate system. Despite the limitations of both methods, their yielding of similar results lends credibility to the estimated shock normal.

\subsection{Shock geometry}

We can determine the shock angle, \( \theta \), which is the angle between the shock normal \( \hat{\textbf{n}} \) and the upstream magnetic field \( \mathbf{B}_{\mathrm{u}} \), using the following definition:

\[ \theta_{Bn} = \arccos \left( \frac{\mathbf{B}_{\mathrm{u}} \cdot \hat{\textbf{n}}}{\|\mathbf{B}_{\mathrm{u}}\| \|\hat{\textbf{n}}\|} \right) . \]

Our calculations yielded shock angles of approximately \( \theta_{Bn}^{\mathrm{MVA}} \sim 12.5^{\circ} \) for the MVA method and \( \theta_{Bn}^{\mathrm{MCT}} \sim 3.9^{\circ} \) for the MCT method. Hence, both methods suggest that the shock was quasi-parallel.

\subsection{Shock speed} \label{Sec:speed}

\begin{figure*}[ht]
    \centering
    \includegraphics[width=1\textwidth]{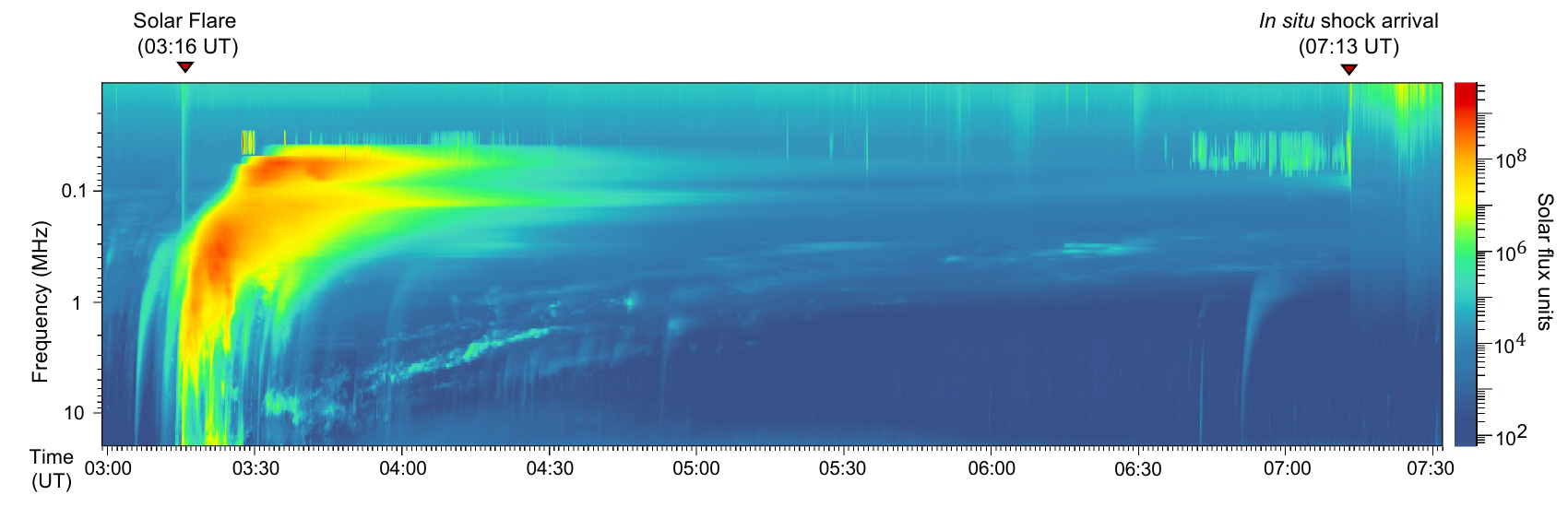}
    \caption{\textbf{Dynamic radio spectra constructed from the FIELDS/RFS receivers.} The Solar flare onset, and the \textit{in situ} shock arrival are indicated on top of the panel.}
    \label{fig:figs2}
\end{figure*}

The shock was measured \textit{in situ} at a distance of approximately 0.23 astronomical units (a.u.), roughly 3 hours and 55 minutes following the solar eruption. The FIELDS/RFS measurements presented in Fig.~\ref{fig:figs2}, the only remote sensing observation in this period, served as the basis for estimating the transit speed. The solar eruption is marked by the dispersionless signal in the FIELDS/RFS receiver when the probe was at 0.24 a.u. This timing suggests a ballistic speed for the eruption-driven shock of approximately 2450~km~s\(^{-1}\). While this offers a basic first-order estimate of the expected speed of the piston (CME), it does not provide an accurate measure of the local shock speed.

Knowing the shock normal vector \(\hat{\textbf{n}}\) is essential for estimating the local speed of the shock, a critical parameter for analyzing different shock rest frames. We utilize the upstream velocity (\(v_{\mathrm{u}} = 410\pm40\)~km~s\(^{-1}\)) and the downstream velocity (\(v_{\mathrm{d}} = 2200 \pm 200\)~km~s\(^{-1}\), along with the shock normal and the expected density compression ratio of a strong shock (\(r_{\mathrm{gas}} \sim 4\)), to calculate the shock speed. The caveat of not having proper plasma measurements downstream is that, it is not possible to obtain the vector direction therefore, only the magnitude of \(v_{\mathrm{d}}\) is used. To estimate the density compression ratio, we use the proxy measurements of density from the FIELDS/RFS: \(n_{\mathrm{d}} = 184 \pm 35\) cm\(^{-3}\) for downstream density and \(n_{\mathrm{u}} = 46.5 \pm15\) cm\(^{-3}\) for upstream density. From these values, we derive the density compression ratio \(r_{\mathrm{gas}} = n_{\mathrm{d}}/n_{\mathrm{u}}\), which approximates to \(4\), aligning with the characteristics of a strong shock. We then use the conservation of mass flux across the parallel shock to estimate the shock speed (\(v_{\mathrm{sh}}\)) as,

\[v_{\mathrm{sh}} = \frac{ (\textbf{v}_{\mathrm{d}} - r_{\mathrm{gas}}^{-1}\textbf{v}_{\mathrm{u}}) \cdot \hat{\textbf{n}}}{1-r_{\mathrm{gas}}^{-1}}\]

\noindent yielding an estimated speed of \(v_{\mathrm{sh}}\sim2800 \pm 300\)~km~s\(^{-1}\) which seems to align with the ballistic estimation. The direction of the vector dot product is completely artificial as only the magnitude of the downstream flow (\(v_{\mathrm{d}}\)) is used in the calculation. Then we estimate the Alfv\'enic Mach number (\(M_{\mathrm{A}}\)) as the ratio of the upstream velocity component normal to the shock (\(\textbf{V}_{\mathrm{u}}\)) to the Alfv\'en speed (\(v_{\mathrm{A}}\)). The Alfv\'en speed is calculated as \(v_{\mathrm{A}} = B_{\mathrm{u}}/\sqrt{\mu_0 n_{\mathrm{u}} m_{\mathrm{p}}}\), where \(B_{\mathrm{u}}\) is the upstream magnetic field magnitude, and \(n_{\mathrm{u}}\) is the upstream proton number density. Here, we use the electron number density obtained from proxies. We then compute \(\textbf{V}_{\mathrm{u}} = \textbf{v}_{\mathrm{u}} \cdot \hat{\textbf{n}} - v_{\mathrm{sh}} \) which results in \(\textbf{V}_{\mathrm{u}} \sim 2400 \pm 300\)~km~s\(^{-1}\). Consequently, for this range of \(\textbf{V}_{\mathrm{u}}\), \(M_{\mathrm{A}}\) is estimated to be approximately \(9.1 \pm 1.35\).

This estimation can be further validated using the proxies established in \citep{Gedalin21}. We can estimate the value of \(M_\mathrm{A}\) using the expression:

\[
\frac{B_{\mathrm{max}}}{B_{\mathrm{u}}} = \sqrt{2M_{\mathrm{A}}^2 (1 - \sqrt{1 - s})+1} .
\]

Here, \( B_{\mathrm{max}} = 330\)~nT and \(B_{\mathrm{u}} = 90\pm5\)~nT are the overshoot and the upstream magnetic field, respectively. The normalized potential jump (\(s\)) is calculated as \(s = 2\phi_{\mathrm{NIF}} / m_{\mathrm{p}} v_{\mathrm{u}}^2\), where \(\phi_{\mathrm{NIF}}\) is the cross-shock electrostatic potential, \( m_{\mathrm{p}} \) is the proton mass, and \( v_{\mathrm{u}} \) is the upstream velocity. The value for \( s = 0.15 \) is selected as a reasonable adjustment of the median cross-shock potential to the magnetic overshoot. This selection is justified based on \cite{Dimmock12}, who demonstrate that for \(M_\mathrm{A} \sim 9\), \(s\) is approximately 0.25, and this value decreases with decreasing \(\theta_{Bn}\). Therefore, a value of \(s = 0.15\) for a nearly parallel shock is justified. Inserting these values into the equation yields \( M_{\mathrm{A}} \sim 9\), which closely aligns with the value estimated from observations.

In a magnetized medium with \(\beta \ll 1\), the critical Mach number (\(M_{\mathrm{c}}\)) for near-parallel shocks can be as low as \(\sim 1.5\) \cite{Kennel85}. The ratio \(M_{\mathrm{A}}/M_{\mathrm{c}} \sim 6.07 \pm 0.9\) clearly indicates that the shock is significantly supercritical. However, energetic particles and associated waves may lead to a change in \(M_{\mathrm{c}}\) \citep{Laming22}

\begin{figure*}[ht]
    \centerline{
    \includegraphics[width=1\textwidth]{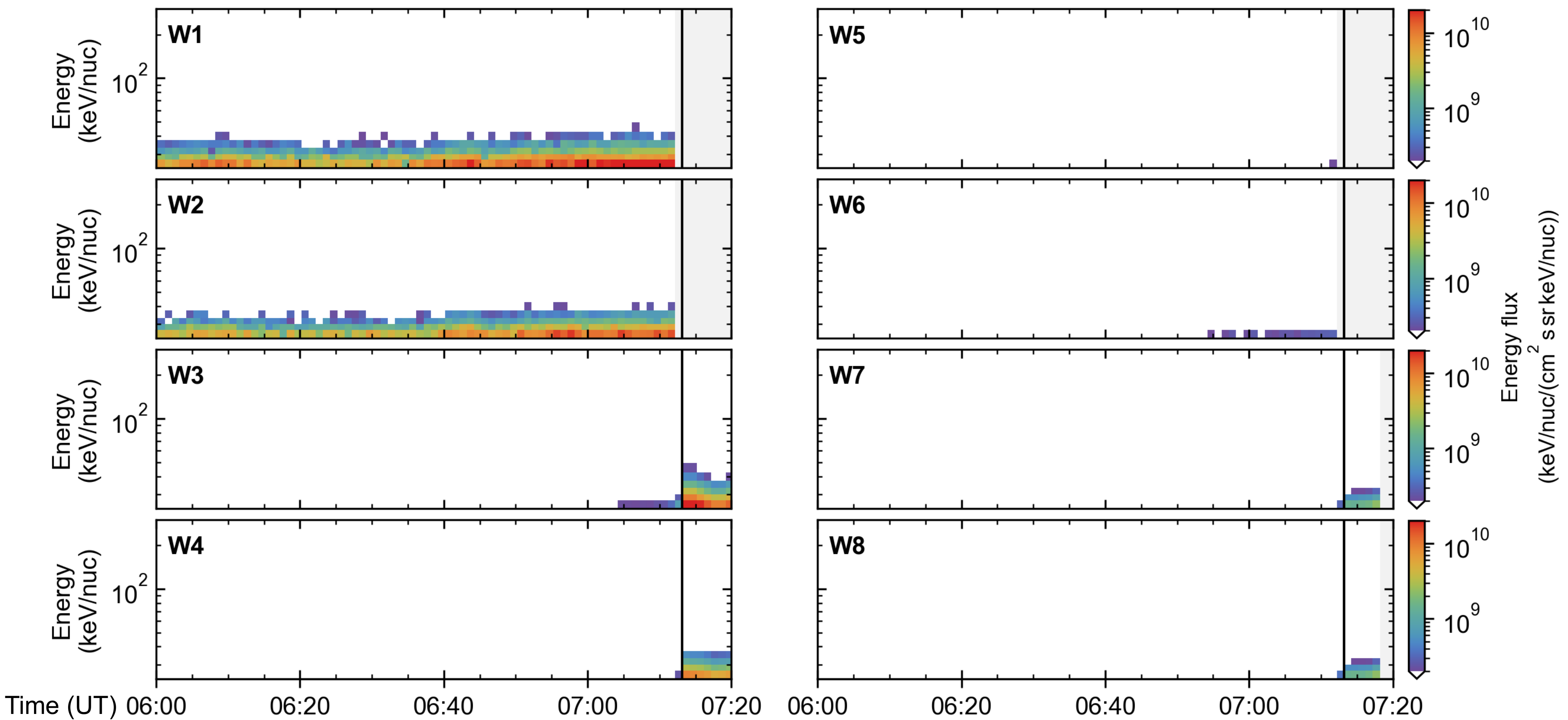}}
    \caption{\textbf{Ion Energy Fluxes Measured by EPI-Lo ChanT, the full time-of-flight data.} The vertical line indicates the moment of the \textit{in situ} shock's arrival. Grey-shaded areas in the graph represent intervals where data was not available.}
    \label{fig:figs3}
\end{figure*}

\begin{figure*}[ht]
    \centerline{
    \includegraphics[width=1\textwidth]{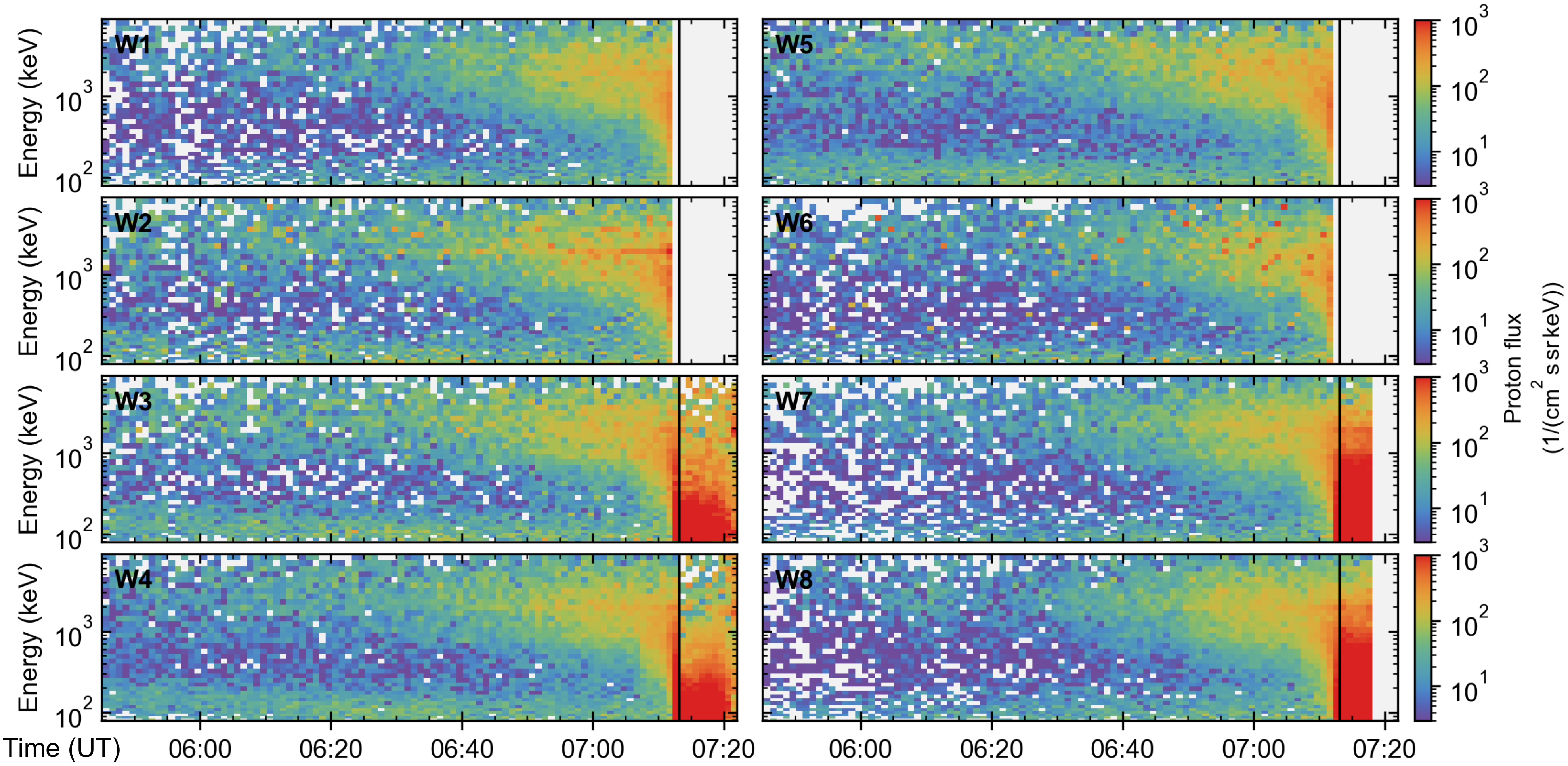}}
    \caption{\textbf{Ion Energy Fluxes Measured by EPI-Lo ChanP, the high energy resolution channel.} The vertical line indicates the moment of the \textit{in situ} shock's arrival. Grey-shaded areas in the graph represent intervals where data was not available.}
    \label{fig:figs4}
\end{figure*}

\begin{figure*}[ht]
    \centerline{
    \includegraphics[width=1\textwidth]{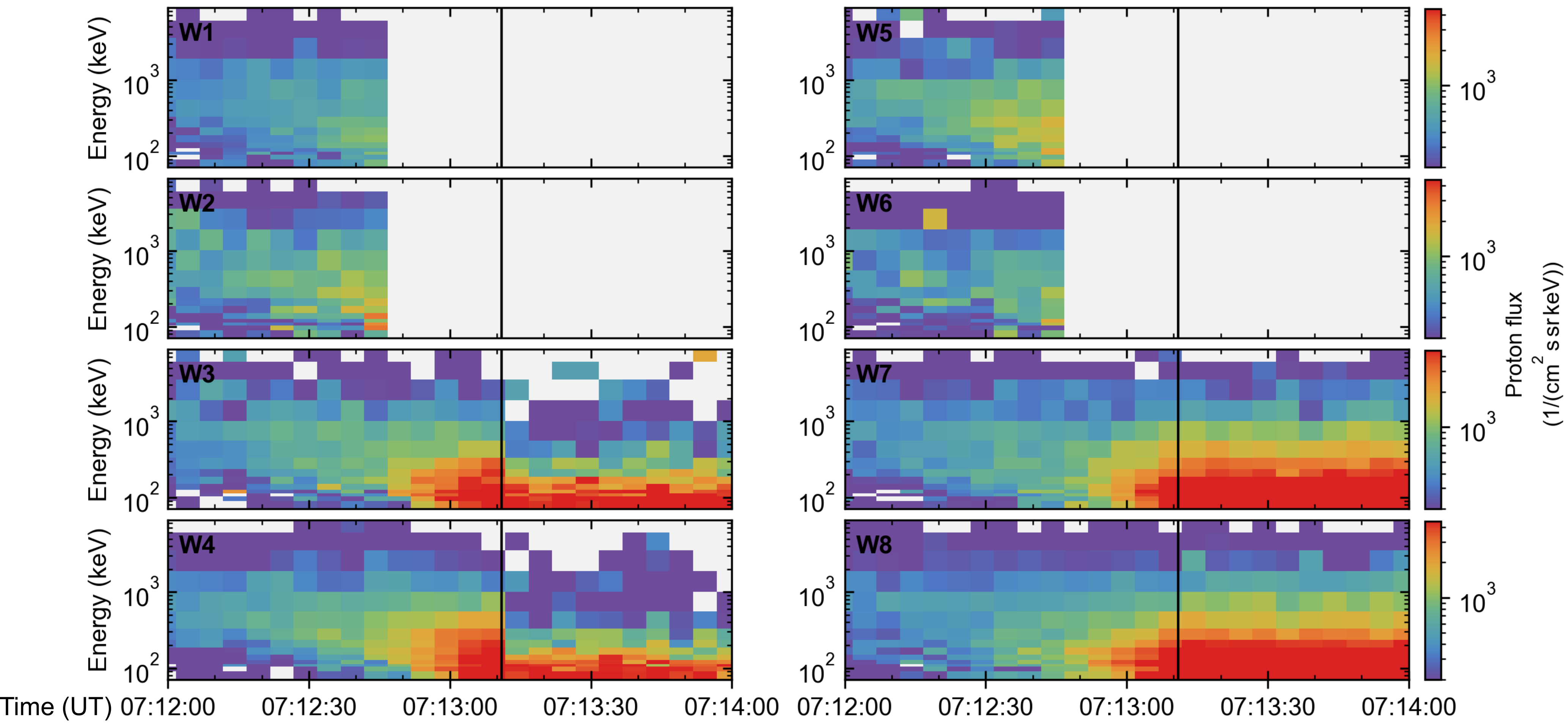}}
    \caption{\textbf{Ion Energy Fluxes Measured by EPI-Lo ChanR, the high time resolution channel.} The vertical line indicates the moment of the \textit{in situ} shock's arrival. Grey-shaded areas in the graph represent intervals where data was not available.}
    \label{fig:figs5}
\end{figure*}

\section{Energetic particle observations} \label{appendix_Epiplots}

During the event, and particularly at the arrival of the interplanetary (IP) shock, particle intensities increased dramatically. Due to these high intensities, EPI-Hi transitioned to dynamic threshold (DT) modes, which involve raising the energy thresholds on various detector segments to limit the effective instrumental geometry factor for incident protons, helium, and electrons. EPI-Lo, which was less affected by the high intensities, was therefore the primary observing instrument.

In response to these conditions, we utilized all available observations from EPI-Lo. Initially, we employed EPI-Lo's full time-of-flight data (ChanT), as shown in Fig.~\ref{fig:fig1}D of the paper, to demonstrate that the bulk flow reached energies within EPI-Lo's range. As ChanT only utilizes particle time-of-flight through the instrument, this channel cannot distinguish ion species.  While Fig.~\ref{fig:fig1}D displays measurements from only the sunward detector, we supplemented this data with measurements from all wedges in Fig.~\ref{fig:figs3}.

Furthermore, we used the energy resolution of EPI-Lo ChanP and the temporal resolution of ChanR to analyze the various proton populations at the shock. Measurements from the sunward pointing detector (W3) are presented in Fig.~\ref{fig:fig2}C of the main text. Complementary measurements from all wedges, supporting the discussions in the main text, are provided in Figs.~\ref{fig:figs4} and \ref{fig:figs5}.

\section{Wave analysis of the upstream} \label{appendix_e}

\subsection{Estimating magnetic field fluctuations} \label{appendix_deltab}
The level of magnetic fluctuations is measured with $\delta B/B$, where $\delta B = |\mathbf{B}-\langle\mathbf{B}\rangle| = |\delta \mathbf{B}|$ and $B = \langle |\mathbf{B}|\rangle$. $\langle\mathbf{B}\rangle$ and $\langle |\mathbf{B}|\rangle$ are calculated as running means of the magnetic field vector and magnitude, respectively, centered around the timestamp. The length of the averaging window $\tau$ is denoted in the plot, e.g., $\left(\delta B/B\right)_{\tau=1\,\mathrm{min}}$. Fluctuations parallel and perpendicular to the mean field are quantified with $\delta B_{\parallel}/B$ and $\delta B_{\perp}/B$, where $\delta B_{\parallel} = |\delta \mathbf{B_{\parallel}}| = |\delta \mathbf{B} \cdot \langle\mathbf{B}\rangle/|\langle\mathbf{B}\rangle||$ and $\delta B_{\perp} = |\delta \mathbf{B_{\perp}}| = \sqrt{\delta B^2_{\perp1}+\delta B^2_{\perp2}}$.

\subsection{Fourier analysis} \label{appendix_FFT}
\begin{figure*}[ht]
    \centerline{
    \includegraphics[width=1\textwidth]{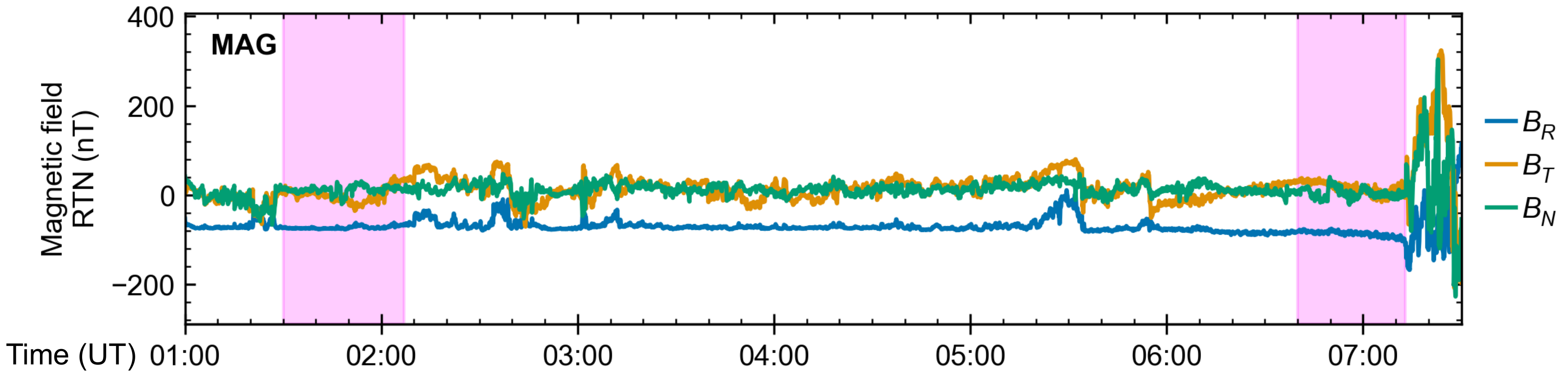}}
    \caption{\textbf{Window selection for Fourier analysis.} The pink shaded areas indicate the FFT (Fast Fourier Transform) windows utilized for estimating the power spectral density shown in Fig.~2F.}
    \label{fig:figs6}
\end{figure*}

In Fig.~2F, we present power spectral densities of the magnetic field in the window 06:40--07:13 UT and in a background window 01:30--02:07. We note that different periods prior to the solar eruption yielded similar background spectra. The background window was chosen to be before the solar eruption at 03:16~UT. This choice is because of the probe's proximity to the Sun, meaning that the fastest particles arrive almost instantaneously in its vicinity. To avoid any potential modifications of the magnetic field influencing our measurements, we selected the pre-event background. Figure~\ref{fig:figs6} shows magnetic field observation during these periods.  We performed a Fourier transform (applying a Hann window) of each magnetic field component $B_R$, $B_T$, and $B_N$. The magnetic field in both intervals is strongly dominated by $B_R$. Therefore, we show the $B_R$ power spectral density, which estimates the power parallel to the background magnetic field, and the sum of $B_T$ and $B_N$ power spectral densities, which estimates the total transverse power. The presented spectra have been smoothed by averaging over adjacent frequencies.

There are two important considerations regarding the data analysis conducted here. The first concerns the window sizes, set at approximately 35 minutes, which correspond to a maximum possible wave period of 2100 seconds (approximately \(5 \times 10^{-4}\) Hz). A wave with a period of 300 seconds (approximately \(3 \times 10^{-3}\) Hz) would be sampled around seven times. Therefore, for frequencies below \(10^{-2}\) Hz or wave periods shorter than 100 seconds (sampled about 21 times), the statistical representation might not accurately reflect reality. The second consideration pertains to the time series analysis itself, which represents a random process—in this case, the magnetic field. The estimations made here are under the assumption of ergodicity, and long time averaging together with the azimuthal motion of PSP may break this behavior. This limitation becomes more significant when PSP is closer to the Sun, affecting the range of wave periods from the same plasma that can be analyzed using Fourier analysis.

\subsection{Estimation of characteristic frequencies} \label{appendix_charac_freq}

In order to understand the effect of the upstream magnetic fluctuations on the ions, we estimate the frequencies of Alfv\'en waves that resonate with 1\,MeV and 5\,MeV protons. When considering waves moving parallel to the background magnetic field, wave-particle cyclotron resonance is governed by the Doppler condition \citep{Vainio00}, 

\[ \omega - k_{\parallel}v_{\parallel} = n\Omega, \]
where $\omega$ is the resonant wave frequency, $k_\parallel$ is the wave number, $v_\parallel$ is the particle speed along the magnetic field, $\Omega$ is the proton cyclotron frequency, and $n$ is an integer. We consider particles with pitch-angle cosine $\mu = 1$, and thus $v_\parallel= v\mu = v$, where $v$ is the speed of the particle. We take $n=1$ corresponding to the fundamental resonance and consider outward-propagating Alfv\'en waves with the dispersion relation $\omega = k_{\parallel}(v_\mathrm{u} + v_\mathrm{A})$ (in the spacecraft frame), where $v_\mathrm{u}$ is the solar wind speed and $v_\mathrm{A}$ is the Alfv\'en speed. Here, we ignore the motion of the spacecraft and approximate that the solar wind flows along the magnetic field. Finally, the resonant frequency becomes

\[ \omega = \frac{v_\mathrm{u}+v_\mathrm{A}}{v_\mathrm{u}+v_\mathrm{A}-v_\parallel}\Omega.\]

\section{Wave analysis of the shock ramp} \label{appendix_f}

\subsection{Morlet wavelet analysis} \label{appendix_morlet}

To determine the properties of the electromagnetic waves in our observations, we applied a Morlet wavelet transform on the time series of both magnetic and electric fields. Wavelets are generally preferred for non-stationary signals such as whistlers and therefore it allows the estimation of power distribution as a function of time and frequency. In short, it reveals the temporal evolution of spectral parameters of wave activity \citep{Bendat11}. Here, we have followed the same wavelet parameters and methodology as used by \cite{Sundkvist12} and \cite{Karbashewski23}.

\subsection{Whistler wave analysis} \label{appendix_whistlers}

An important aspect of any quasi-parallel shock is the number of large amplitude magnetic structures throughout the magnetic gradient of the shock. It is widely understood that the general structure of the quasi-parallel shock is highly non-stationary and therefore time-dependent. At a shock propagating at high speeds such as the one studied here, we would require a close to 1~kHz resolution for both magnetic fields and plasma properties to truly understand the evolution in detail. In our observations, the FIELDS instrument suite was the only one capable of capturing phenomena in the scale of several ion-inertial lengths ($\lambda_\mathrm{i} = c/\omega_{\mathrm{pi}}$). We performed a spectral and phase wave analysis of the data which revealed several features in the vicinity the the shock. The large amplitude waves in the frequency range of 0.1--10 Hz are located close to the ramp, both upstream and downstream. The observed local peak of the electric field indicates the crossing of the ramp at 07:13:11.45. The ion-inertial length estimation is based on electron density processing using the FIELDS plasma waves measurement. In the upstream, we found the $n_{\mathrm{e}}$ to be $\sim 46.5 \pm 15$ cm$^{-3}$. Assuming that the plasma is quasi-neutral, the ion density may not vary far from the electron density, and as such would correspond to an ion-inertial length, $\lambda_\mathrm{i} \sim 32.5 \pm 3$ km. 


The distinct whistler precursor is observed in the upstream region of the quasi-perpendicular ramp. Several wave periods are seen at frequencies around 30~Hz. The waves are right-hand circularly polarized (Fig.~\ref{fig:fig3}H) and have amplitudes up to 35~nT. The wave normal angles (WNA, \(\theta_{kB}\)) are derived by making use of the singular value decomposition (SVD) technique \citep{Karbashewski23}. The WNA for the whistler wave was found to be around $73\pm3^{\circ}$ (Fig.~\ref{fig:fig3}E). The oblique propagation is the essential factor for the wave precursor to be a standing wave in the shock frame. 

The Poynting flux can only reliably be determined along the Sun--spacecraft $\mathrm{R}$-axis due to the unreliable $\mathrm{R}$-component electric field measurement. The Poynting flux in the spacecraft frame is calculated using the complex $\vec{E}_w$ and $\vec{B}_w$ spectra: $S_z = Re(E_{wx}\cdot B_{wy}^*-E_{wy}\cdot B_{wx}^*)$ from Fig.~\ref{fig:fig3}(C,D) and subsequently mapped to a bi-symmetric logarithmic scale shown in Fig.~\ref{fig:fig3}F. The wave phase speed (\(v_{\mathrm{ph}}\)) is estimated using the flow speed and the WNA. 

\subsubsection{Schematic of the Shock transition}
The schematic in Fig.~\ref{fig:fig4}B presents a 2D simplified reconstruction of the shock structure based on the values and direction of the magnetic field recorded by PSP during the shock crossing presented in Fig.~\ref{fig:fig4}C. The changing velocity of the plasma flow is reflected in the nonlinear (but isotropic in each particular point of the schematic) spatial scales. The physical scales are indicated by the squares of $50 \lambda_{\mathrm{i}}~\times~50 \lambda_{\mathrm{i}}$. The color scheme represents the magnetic field magnitude.

\bibliography{bibTeX_11_11_2022}{}
\bibliographystyle{aasjournal}

\end{document}